\documentclass[twocolumn]{aastex631}

\usepackage{amsmath}
\usepackage[utf8]{inputenc}
\usepackage{xcolor}
\usepackage{graphicx}
\usepackage{tikz}

\usepackage{booktabs} % For nicer lines in tables.

\usepackage{svg} % For nicer lines in tables.
\usepackage{float}

\usepackage{bm}
\usepackage{soul}

\defcitealias{CH09}{CH09}
\defcitealias{Cr09}{Cr09}
\defcitealias{PP98}{PP98}
\defcitealias{fox_solar_2016}{PSP}
\defcitealias{muller_solar_2020}{SolO}

\defcitealias{van1995python}{\texttt{Python}}
\defcitealias{2020SciPy-NMeth}{\texttt{SciPy}}
\defcitealias{mckinney2010data}{\texttt{Pandas}}
\defcitealias{Hunter2007Matplotlib}{\texttt{Matplotlib}}
\defcitealias{PYMC_2015}{\texttt{PyMC}}
\defcitealias{angelopoulos_space_2019}{\texttt{Pyspedas}}
\defcitealias{MHDTurbPy_Sioulas}{\texttt{MHDTurbPy}}

\definecolor{airforceblue}{rgb}{0.36, 0.54, 0.66}
\definecolor{bluegray}{rgb}{0.4, 0.6, 0.8}

\usepackage{hyperref, array}
\usepackage{CJK}

\begin{document}

\shortauthors{Sioulas et al.}

\correspondingauthor{Nikos Sioulas}
\email{nsioulas@g.ucla.edu}

\begin{CJK*}{UTF8}{gbsn}

\title{On the Propagation and Damping of Alfvenic Fluctuations in the Outer Solar Corona \& Solar Wind}

\author[0000-0002-1128-9685]{Nikos Sioulas}
\affiliation{Space Sciences Laboratory, University of California, Berkeley, CA 94720-7450, USA}

\author[0000-0002-2381-3106]{Marco Velli}
\affiliation{Department of Earth, Planetary, and Space Sciences, University of California, Los Angeles, CA, USA}

\author[0000-0002-2582-7085]{Chen Shi (时辰)}
\affiliation{Department of Earth, Planetary, and Space Sciences, University of California, Los Angeles, CA, USA}

\author[0000-0002-4625-3332]{Trevor A. Bowen}
\affiliation{Space Sciences Laboratory, University of California, Berkeley, CA 94720-7450, USA}

\author[0000-0003-4177-3328]{Alfred Mallet}
\affiliation{Space Sciences Laboratory, University of California, Berkeley, CA 94720-7450, USA}

\author[0000-0003-4380-4837]{Andrea Verdini}
\affiliation{Universit\'a di Firenze, Dipartimento di Fisica e Astronomia, Firenze, Italy}

\author[0000-0003-4177-3328]{B. D. G. Chandran}
\affiliation{Space Science Center and Department of Physics, University of New Hampshire, Durham, NH 03824, USA}

\author[0000-0003-2880-6084]{Anna Tenerani}
\affiliation{Department of Physics, The University of Texas at Austin, TX 78712, USA}

\author[0000-0002-1628-0276]{Jean-Baptiste Dakeyo}
\affiliation{Space Sciences Laboratory, University of California, Berkeley, CA 94720-7450, USA}

\author[0000-0002-1989-3596]{Stuart D. Bale}
\affil{Space Sciences Laboratory, University of California, Berkeley, CA 94720-7450, USA}
\affil{Physics Department, University of California, Berkeley, CA 94720-7300, USA}

\author[0000-0001-5030-6030]{Davin Larson}
\affiliation{Space Sciences Laboratory, University of California, Berkeley, CA 94720-7450, USA}

% \author[0000-0001-8479-962X]{Jonathan Squire}
% \affiliation{Physics Department, University of Otago, Dunedin 9010, New Zealand}

\author[0000-0001-5258-6128]{Jasper S. Halekas}
\affiliation{Department of Physics and Astronomy, University of Iowa\\
Iowa City, IA 52242, USA}

\author[0000-0002-6276-7771]{Lorenzo Matteini}
\affiliation{Imperial College London, South Kensington Campus, London SW7 2AZ, UK}

\author[0000-0002-2916-3837]{Victor R\'eville}
\affiliation{IRAP, Universit\'e de Toulouse,
CNRS, CNES, Toulouse, France}

\author[0000-0003-4529-3620]{C. H. K. Chen}
\affiliation{Department of Physics and Astronomy, Queen Mary University of London, London E1 4NS, UK}

\author[0000-0002-4559-2199]{Orlando M. Romeo}
\affil{Space Sciences Laboratory, University of California, Berkeley, CA 94720-7450, USA}

\author[0000-0003-2981-0544]{Mingzhe Liu}
\affil{Space Sciences Laboratory, University of California, Berkeley, CA 94720-7450, USA}

\author[0000-0002-0396-0547]{Roberto Livi}
\affiliation{Space Sciences Laboratory, University of California, Berkeley, CA 94720-7450, USA}

\author[0000-0003-0519-6498]{Ali Rahmati}
\affiliation{Space Sciences Laboratory, University of California, Berkeley, CA 94720, USA.}

\author[0000-0002-7287-5098]{P. L. Whittlesey}
\affiliation{Space Sciences Laboratory, University of California, Berkeley, CA 94720, USA.}

\begin{abstract}

We analyze \textit{Parker Solar Probe} and \textit{Solar Orbiter} observations to investigate the propagation and dissipation of Alfv\'enic fluctuations from the outer corona to 1~AU. Conservation of wave-action flux provides the theoretical baseline for how fluctuation amplitudes scale with the Alfv\'en Mach number $M_a$, once solar-wind acceleration is accounted for. Departures from this scaling quantify the net balance between energy injection and dissipation. Fluctuation amplitudes follow wave-action conservation for $M_a < M_a^{b}$ but steepen beyond this break point, which typically lies near the Alfv\'en surface ($M_a \approx 1$) yet varies systematically with normalized cross helicity $\sigma_c$ and fluctuation scale. In slow, quasi-balanced streams, the transition occurs at $M_a \lesssim 1$; in fast, imbalanced wind, WKB-like scaling persists to $M_a \gtrsim 1$. Outer-scale fluctuations maintain wave-action conservation to larger $M_a$ than inertial-range modes. The turbulent heating rate $Q$ is largest below $M_a^{b}$, indicating a preferential heating zone shaped by the degree of imbalance. Despite this, the Alfv\'enic energy flux $F_a$ remains elevated, and the corresponding damping length $\Lambda_d = F_a/Q$ remains sufficiently large to permit long-range propagation before appreciable damping occurs. Normalized damping lengths $\Lambda_d/H_A$, where $H_A$ is the inverse Alfv\'en-speed scale height, are near unity for $M_a \lesssim M_a^{b}$ but decline with increasing $M_a$ and decreasing $U$, implying that incompressible reflection-driven turbulence alone cannot account for the observed dissipation. Additional damping mechanisms---such as compressible effects---are likely required to account for the observed heating rates across much of the parameter space.

\end{abstract}

%,  provided the data are binned by solar-wind speed to avoid mixing of streams with different acceleration profiles and velocities

% \keywords{ Solar Wind; Plasmas; Turbulence; Waves}

\section{Introduction}\label{sec:intro} 

The continuous outflow of hot plasma is now recognized as a ubiquitous feature of stars across nearly all spectral types and evolutionary stages on the Hertzsprung-Russell diagram. \citet{parker_dynamics_1958} demonstrated that, in the limit of negligible interstellar pressure, such outflows must accelerate to supersonic speeds beyond a critical radius, establishing the theoretical foundation for the solar wind\footnote{The response of spherically symmetric, non-rotating atmospheres to external conditions can be more complex; see \citet{velli_supersonic_1994}.}.

Early models treated the solar wind as a spherically symmetric, thermally driven outflow from a hot coronal base. Pressure-driven fluid solutions \citep{Parker_1965, Weber_Davis_1967} and related exospheric approaches \citep{SEN_1969, Lemaire_1971, Zouganelis_2004, Parker_2010} successfully reproduced the bulk properties of the slow solar wind \citep[see, e.g.,][]{Halekas_2022, Horaites_2022, 2023_Halekas}. However, these models fail to account for the elevated proton temperatures and high bulk speeds observed in the fast wind near 1~au \citep[see, e.g.,][and references therein]{Leer_1982_review, Hansteen_Velli_2012}---features now widely interpreted as signatures of spatially extended heating and momentum deposition along the flow \citep{Withbroe_1977, Holzer_Leer_1980, Shi_poly_2022, 2023_Halekas}.

One proposed resolution to these discrepancies invokes the presence of large-amplitude Alfv\'en waves (AWs)\footnote{Exact nonlinear solutions of the homogeneous MHD equations in which the fluctuating velocity and magnetic fields satisfy $\boldsymbol{U} = \pm \boldsymbol{B}/\sqrt{4\pi\rho}$, while the magnetic-field strength, density, and thermal pressure are space-time constants ($\delta|\mathbf{B}| = \delta\rho = \delta p = 0$). As a result, AWs dissipate over length scales far greater than those of compressive acoustic or magnetosonic waves \citep[see, e.g.,][]{Barnes_1966_damping}} \citep[see, e.g.,][]{Osterbrock_1961, Belcher_1971_AW_sw_model, Alazraki_1971, Hollweg_1974, perez_how_2021}. AWs can both exert (in a time-averaged sense) a net outward force on the plasma and supply the heating required to accelerate the solar wind to its observed terminal speeds and to account for its gradual, non-adiabatic cooling at and beyond 1~au \citep[e.g.,][]{hellinger_heating_2011, Bowen_2025}.

The first \textit{in~situ} spacecraft measurements of large-amplitude, outward-propagating Alfv\'enic fluctuations \citep{Unti_Neugebauer_1968, Belcher_1969} provided compelling empirical support for this hypothesis. More recently, remote-sensing observations have identified low-frequency Alfv\'enic motions in the lower solar corona, carrying sufficient energy and momentum to heat the plasma and drive the solar wind \citep{2007_de_Pontieu, tomczyk_2007, Morton_2015}.

It is now widely recognized that a comprehensive understanding of the coupled problem of coronal heating and mechanically driven stellar winds demands detailed knowledge of how Alfv\'enic fluctuations are transported through, and interact with, an inhomogeneous, radially expanding plasma \citep[see, e.g.,][and references therein]{Matthaeus11, bruno_solar_2013}.

In the solar wind, the bulk of the fluctuating energy resides at scales much larger than the characteristic ion kinetic scales,\footnote{Specifically, well above the ion gyroradius and inertial length, defined as $\rho_i = V_{\text{th},i}/\Omega_i$ and $d_i = V_A/\Omega_i$, respectively, where $V_{\text{th},i} = \sqrt{k_B T_i / m_p}$ is the ion thermal speed, and $\Omega_i = e |\mathbf{B}| / m_p$ is the ion cyclotron frequency.} thereby justifying the use of a magnetohydrodynamic (MHD) framework to describe AW dynamics \citep{Belcher_Davis_1971}. The governing equations for incompressible fluctuations in an inhomogeneous, stationary, magnetized plasma are commonly expressed in terms of the \citet{elsasser_1950} variables, $\bm{z}_{\pm} = \bm{v} \mp \text{sign}(\bm{B}_0)\bm{b} / \sqrt{4\pi\rho}$, where $\bm{B}_0$ and $\bm{b}$ denote the large-scale and fluctuating magnetic fields, respectively, and $\bm{v}$ represents fluctuations in the plasma velocity \citep{Velli_1990, Velli_93}:

\onecolumngrid
\begin{tikzpicture}[remember picture,overlay]
  \draw (0,0) -- (8.5,0); % horizontal line at the top
  \draw (0,0) -- (0,0); % vertical line left
 \draw (8.5,0) -- (8.5,0.5); % vertical line right
\end{tikzpicture}

\begin{equation}\label{eq:MHD}
\frac{\partial \boldsymbol{z}_{\pm}}{\partial t} + (\boldsymbol{U} \pm \boldsymbol{V_A}) \cdot \nabla \boldsymbol{z}_{\pm} + \boldsymbol{z}_{\mp} \cdot \nabla (\boldsymbol{U} \mp \boldsymbol{V_A}) + \frac{1}{2} (\boldsymbol{z}_{-} - \boldsymbol{z}_{+}) \nabla \cdot \left( \boldsymbol{V_A} \mp \frac{1}{2} \boldsymbol{U} \right) = \mathcal{F}_{nl},
\end{equation}

\begin{tikzpicture}[remember picture,overlay]
% Empty line retained for spacing
  \draw (8.5,0) -- (8.5,-0.5); % vertical line right
  \draw (8.5,0) -- (17,0); % horizontal line at the bottom
\end{tikzpicture}
\twocolumngrid

 and $\bm{U}$, $ \bm{V}_A =\bm{B}_{0} / \sqrt{4\pi \rho}$ denote the mean (large-scale), flow and Alfv\'en velocity.
$\mathcal{F}_{nl}$ denotes the non-linear terms $\mathcal{F}_{nl} = -(\rho^{-1} \nabla p_{tot} + \boldsymbol{z}_{\mp} \cdot \nabla \boldsymbol{z}_{\pm})$,
where $p_{tot}$ is the total (kinetic plus magnetic) pressure , determined in the incompressible limit by enforcing $\nabla \cdot \boldsymbol{z}_{\pm} = 0$ through the divergence of the equation of motion.

The first two terms on the RHS of Equation~\ref{eq:MHD} describe wave propagation. The third term accounts for the reflection of finite-wavelength Alfv\'en waves arising from spatial gradients in $U \mp V_{A}$ along the direction of the fluctuations. These gradients couple the oppositely directed Els\"asser modes, $\boldsymbol{z}_{\pm}$, such that in a non-uniform background plasma, a purely outward-propagating wave is no longer an exact solution \citep{Heinemann_Olbert}. The fourth term captures both the isotropic component of the reflection and the WKB\footnote{The term ``WKB limit'' (Wentzel-Kramers-Brillouin) is used broadly to denote an asymptotic expansion method for solving linear differential equations. In this work, it refers specifically to purely outward propagation under idealized conditions---namely, in the absence of both reflection and nonlinear effects.} amplitude decay of the fluctuations \citep{Weinberg_1962, Hollweg_1974}.

A direct consequence of Equation~\ref{eq:MHD} in the linear limit ($\mathcal{F}_{\text{nl}} = 0$), assuming a steady-state background, is that high-frequency, short-wavelength\footnote{That is, relative to the characteristic temporal and spatial scales over which background plasma parameters vary appreciably.} AWs propagate adiabatically, conserving the total wave action flux, $\mathcal{S}^{\ast} = S^{+} - S^{-}$ \citep{Heinemann_Olbert, Velli_93, 2015_Chandran_wave_action}; see Appendix~\ref{Appendix_Theory:Wave_action}. The associated wave stress\footnote{We use the term stress rather than pressure to emphasize that wave-induced forces are generally anisotropic.} \citep{bretherton1968wavetrains, Dewar_1970} does work on the expanding atmosphere and supplies a non-thermal mechanism for accelerating the solar wind \citep{Parker_1965, Alazraki_1971, Belcher_1971_AW_sw_model}.

\par In the limit of negligible reflection ($z_{+} \gg z_{-}$, implying $S^{-} \approx 0$), wave-action conservation reduces to $S^{+} = \text{const.}$ This corresponds to the WKB limit \citep{witham1965general} of non-interacting AWs streaming away from the Sun \citep{Parker_1965, Jacques_1977}.

After traversing the lower solar atmosphere and the transition region---where the steep Alfv\'en-speed gradient acts as a high-pass filter \citep{Leroy_1981}, suppressing the transmission of hour-scale fluctuations \citep{Velli_93, Reville_2018, Chandran_Perez_2019}---AWs enter the corona. There, they propagate outward and are expected to undergo partial, non-WKB reflection, predominantly at frequencies small compared to the expansion rate. A frequency-dependent transmission coefficient, defined as the ratio of wave-action fluxes in inward- and outward-propagating wave packets, can thus be used to characterize their propagation \citep[see, e.g.,][]{Velli_91_waves, Charbonneau_1995_Non_WKB}.

The observed spectral peak in the near-Sun solar wind---manifested as a shallower-than-$1/f$ slope at low frequencies \citep{Tu95, kasper_PSP_alfven, Davis_2023, 2023_Huang}---suggests that the large-scale fluctuations dominating the inner-heliospheric spectrum must, at least in part, develop in situ, implying the presence of an inverse transfer of energy. 

Such inverse transfer is permitted within the framework of incompressible, inhomogeneous MHD turbulence \citep{velli_turbulent_1989, Perez_Chandran_2013}, and has recently been interpreted by \citet{meyrand2023reflectiondriven} as a consequence of anomalous growth in wave-action anastrophy under reflection-driven dynamics. However, compressible processes may also play a significant role. In particular, the parametric decay instability (PDI)---in which large-amplitude, outward-propagating Alfv\'en waves nonlinearly couple to slow magnetosonic modes \citep{Galeev_1963, Goldstein_1978, 1997_prunareti_PDI, Reville_2018, Malara_22_PDI}---produces two daughter waves: a sunward-directed Alfv\'en wave with slightly reduced frequency and an antisunward-propagating slow mode. In this way, PDI facilitates an inverse transfer that populates the low-frequency spectrum \citep{chandran_2018, Reville_2018}, while simultaneously generating strong density perturbations that can steepen into shocks and substantially enhance dissipation rates relative to those expected under purely incompressible dynamics \citep{vanBallegooijen_2016, 2019ApJ_Shoda}.

Collisions between the reflected and outward-propagating wave packets drive nonlinear couplings \citep{iroshnikov_turbulence_1963, kraichnan_inertial-range_1965}, giving rise to a strongly anisotropic cascade\footnote{In recent years, reflection-driven AW turbulence has emerged as a leading framework for explaining how AWs energize the solar wind \citep{Zhou_Matthaeus_1989, velli_turbulent_1989, Dmitruk_2002, Cranmer_2005, Verdini_2007, Verdini_2009ApJ, Perez_Chandran_2013, Chandran_Perez_2019, meyrand2023reflectiondriven}.
} in which energy is preferentially transferred to smaller scales perpendicular to the background magnetic field, $\bm{B}_0$ \citep{shebalin_matthaeus_montgomery_1983, higdon_anisotropic}. The shear AW cascade extends down to ion kinetic scales, at or below which it terminates through the irreversible conversion of wave energy into heat via collisionless damping. The specific dissipation mechanisms at play determine both the damping scale and how the dissipated energy is partitioned among different plasma species \citep[see, e.g.,][and references therein]{schekochihin_astrophysical_2009, Howes_2024}.

Building on this framework, wave-action flux conservation provides a physically grounded reference against which the radial evolution of fluctuation amplitudes can be assessed. Departures from the WKB-predicted scaling quantify the net effect of energy exchange processes. A shallower-than-expected decay suggests local replenishment of wave energy---for example, through compression or shear associated with stream interactions \citep{Coleman_68, roberts_velocity_1992, Shi_2020, shi_influence_2022}---whereas a steeper decay indicates reflection followed by enhanced nonlinear transfer and dissipation \citep{1968ApJ...153..371C, Hollweg_1986}.

Observations from multiple spacecraft confirm that, in the low-frequency portion of the spectrum, the fluctuation energy decays approximately as $\langle \delta B^2 \rangle \propto R^{-3}$, consistent with the WKB scaling implied by the condition $S^{+} = \text{const.}$ under a set of simplifying assumptions (see Appendices~\ref{Appendix_Theory:Wave_action} and~\ref{Appendix_methods:WKB_adherence}). In contrast, fluctuations at higher frequencies---corresponding to the inertial range---typically exhibit a steeper radial decay \citep{Villante_1982, bavassano_radial_1982, Roberts_1987, Horbury_2001, Tenerani_2021}.

\begin{figure*}
     \centering
        \includegraphics[width=0.99\textwidth]{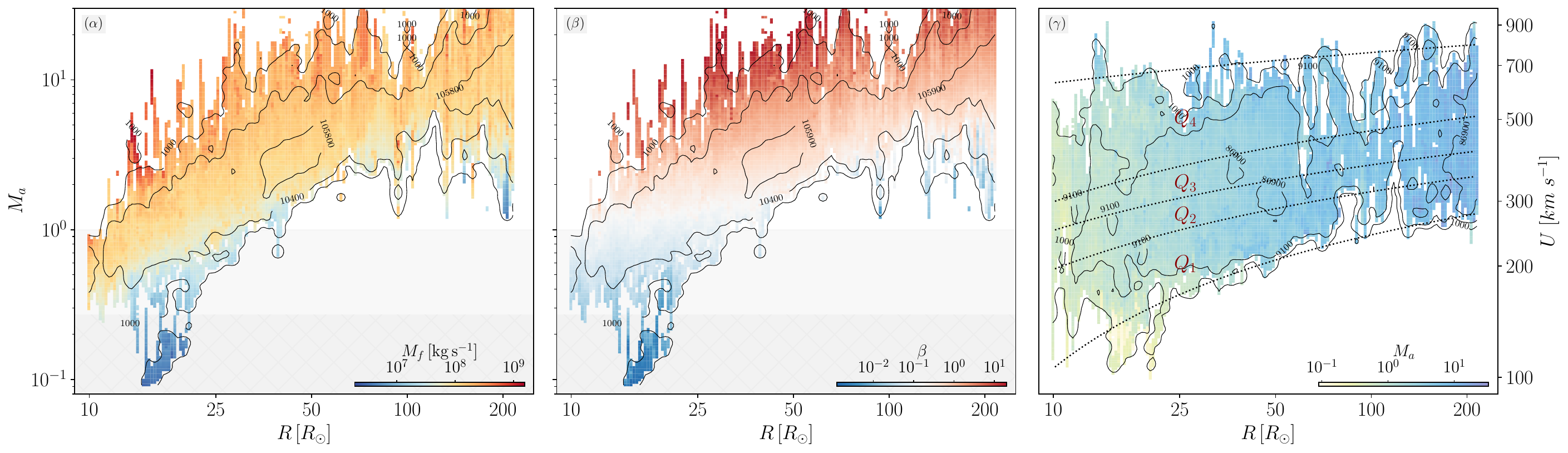}
        \caption{ Two-dimensional distributions of key solar-wind parameters. ($\alpha$) Mass flux, $M_f$, and ($\beta$) plasma $\beta$, $\beta$, plotted as functions of heliocentric distance $R$ (horizontal axis) and Alfv\'en Mach number $M_a$ (vertical axis). The color scale indicates the magnitude of each quantity, while black contour lines trace the data-point density; inline labels mark the number of intervals per bin. ($\gamma$) Alfv\'en Mach number, $M_a$, shown as a function of $R$ and solar-wind speed $U$. Dotted lines correspond to isothermal Parker-wind velocity profiles fitted to quantile boundaries within each of the $N_x = 100$ logarithmically spaced radial bins. In each bin, the data are divided into four equal-sized quantiles, $Q_i(R)$ for $i = 1,\dots,4$, and the fitted profiles $U_i(R)$ serve as radial-dependent thresholds for classifying the solar wind into four distinct acceleration regimes.
}\label{fig:heatmaps}
\end{figure*}

As the solar wind expands radially, it undergoes a fundamental transition at the Alfv\'en surface, $R_A$, defined by the condition $U = V_A = B / \sqrt{\mu_0 \rho}$ \citep[see, e.g.,][]{Cranmer_2023}. This surface separates the magnetically dominated sub-Alfv\'enic corona ($M_A \equiv U/V_A < 1$) from the super-Alfv\'enic outer solar wind ($M_A > 1$), beyond which plasma and magnetic structures are no longer causally connected to the solar surface. Within the WKB framework, the energy density of outward-propagating Alfv\'en waves is expected to reach a maximum near $R_A$ due to the conservation of wave-action flux in an expanding flow \citep[see, e.g.,][]{Tenerani_EBM}; see also Appendix~\ref{Appendix_Theory:Wave_action}.

However, the precise behavior of fluctuation amplitudes near $R_A$ depends sensitively on model assumptions. Some treatments predict a local depletion of wave power \citep{Adhikari_2019}, while others suggest an enhancement associated with large-scale coronal shear \citep{Ruffolo2020}. Turbulence-transport models that incorporate both partial wave reflection and nonlinear cascade predict that the amplitude maximum is established well inside the sub-Alfv\'enic corona, at heliocentric distances smaller than $R_A$ \citep[see, e.g.,][]{Verdini_velli_2007, Verdini_2010, Chandran_Perez_2019}.

The absence of \textit{in~situ} measurements within the sub-Alfv\'enic solar wind long limited direct observational tests of theoretical predictions in this region. This constraint has now been overcome by the \textit{Parker Solar Probe} \citep[PSP;][]{fox_solar_2016}, which has provided the first \textit{in~situ} access to coronal plasma \citep{kasper_PSP_alfven}, capturing the early stages of solar wind formation. In combination with \textit{Solar Orbiter} \citep[SolO;][]{muller_solar_2020}, these missions now enable continuous tracking of Alfv\'enic wave dynamics from the extended corona to 1~AU, offering new constraints on their role in solar wind acceleration and heating \citep{2020_Velli_alignments, Tenerani_2021, 2023_Halekas, Shi_2023_ion_elec, Rivera_2024, Bourouaine_2024, Bowen_2025, Bowen_2025_non_linear}.

\citet{Ruffolo_2024} analyzed the evolution of turbulence amplitudes across the Alfv\'en surface,\footnote{A similar study---comparable in scope but differing in interpretation---was presented at recent conferences \citep{2023AGUFMSH31D3014C}.} reporting an enhancement of wave action within the sub-Alfv\'enic regime. This feature was attributed to energy injection near $R_A$, driven by large-scale coronal shear flows. However, the interpretation relies on several simplifying assumptions and methodological choices that may limit the robustness of the result. These considerations motivate a more cautious reassessment of their conclusions.

\begin{figure*}
     \centering
        \includegraphics[width=0.99\textwidth]{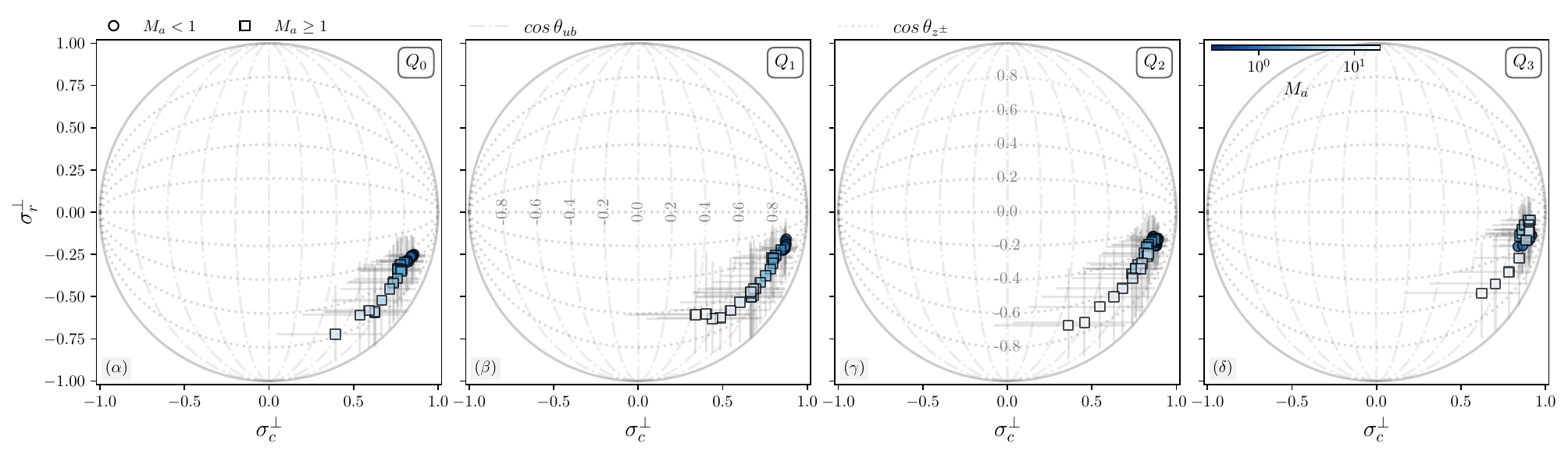}
        \caption{ Parametric representation of $\sigma^{\perp}_r$ and $\sigma^{\perp}_c$ as functions of $M_a$. Each marker denotes the median within one of 30 logarithmically spaced $M_a$ bins, with horizontal and vertical error bars indicating the interquartile range (25th-75th percentiles). Circles are used for bins with $M_a \leq 1$, while squares denote bins with $M_a > 1$. Panels (a-d) correspond to the four quartiles of the solar wind acceleration profiles (see Appendix \ref{Appendix:methods}). The logarithmic color scale encodes the bin-centered value of $M_a$. Horizontal dashed gray lines indicate contours of constant Els\"asser increment alignment angle, defined by $\cos\, \theta_{z^{\pm}} = \sigma^{\perp}_{r}/\sqrt{1 - (\sigma^{\perp}_{c})^{2}}$, while vertical dash-dotted gray lines mark contours of constant velocity-magnetic field increment alignment angle, defined by $\cos\, \theta_{ub} = \sigma^{\perp}_{c}/\sqrt{1 - (\sigma^{\perp}_{r})^{2}}$.}\label{fig:sigmas}
\end{figure*}

% \begin{figure}
% \begin{center}

%         \includegraphics[width=0.34\textwidth]{figures/Cb_vs_Ma.pdf}
%         \caption{Magnetic compressibility, $C_{B}$, as a function  of $M_a$.}\label{fig:Cb_Ma}
%     \end{center}
% \end{figure}

In the following, we analyze an extended dataset comprising multi-encounter observations from both \citetalias{fox_solar_2016} and \citetalias{muller_solar_2020} to investigate the propagation and damping of Alfv\'enic fluctuations from the outer corona into the solar wind. Fluctuation amplitudes are broadly consistent with wave-action conservation up to a characteristic Alfv\'en Mach number, $M_a^b$, beyond which both the wave-action flux and outward-directed power transition to a steeper, power-law decay. While $M_a^b$ typically lies near the Alfv\'en critical point, its precise location varies systematically with solar-wind speed and fluctuation scale. Wave-action conservation persists even though the volumetric heating rate $Q$ peaks at or below $M_a^b$, as the Alfv\'en wave energy flux $F_a$ remains elevated in the same region. The resulting large damping lengths, $\Lambda = F_a/Q$, imply that outward-propagating wavepackets can traverse extended distances before undergoing significant dissipation.

The structure of the paper is as follows. Section~\ref{sec:Statistical Results} presents the main observational results, focusing on the evolution of Alfv\'enic fluctuation amplitudes, turbulent heating rates, and damping lengths as functions of solar-wind speed and Alfv\'en Mach number. Section~\ref{sec:Conclusions} summarizes the key findings and their implications for wave-driven solar-wind models. Appendix~\ref{Appendix:background} provides the theoretical background on wave-action conservation in inhomogeneous flows, including the WKB limit and the expected $M_a(R)$-dependent scaling. It also reviews the turbulence-based heating models of \citet{CH09}, \citet{Cr09}, and \citet{PP98}, emphasizing their respective assumptions and dissipation mechanisms. Appendices~\ref{App:Data_Sel_and_Processing} and~\ref{Appendix:methods} detail the data selection, preprocessing, and analysis methods used throughout.

\begin{figure*}
        \centering
         \includegraphics[width=1\textwidth]{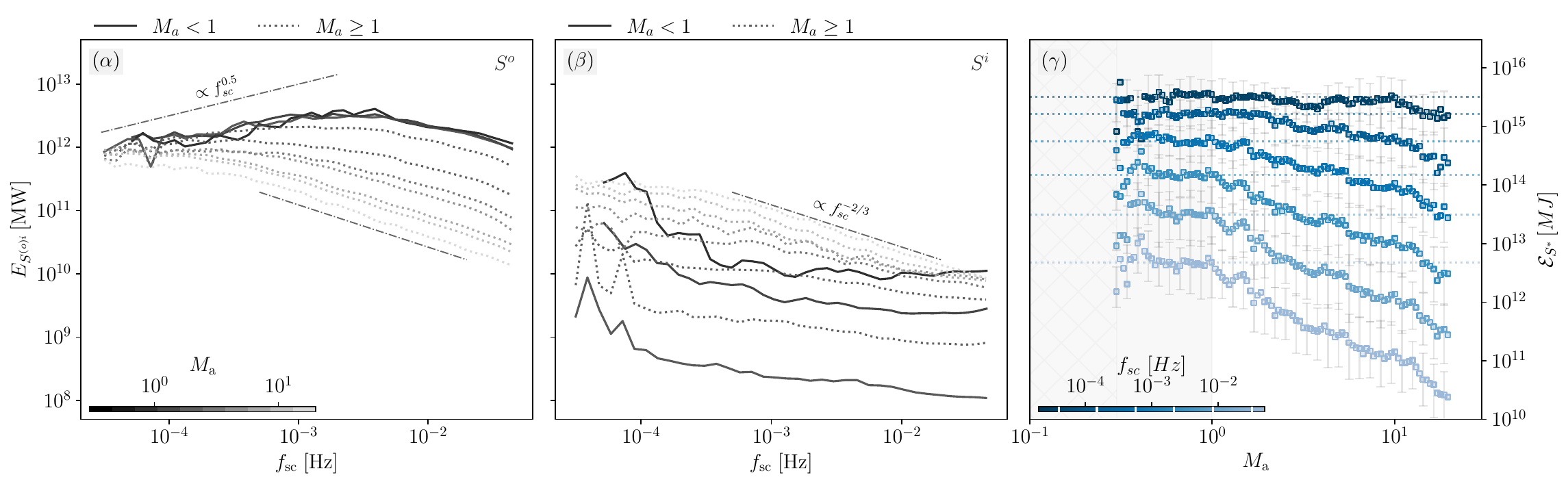}
        
        \caption{Wave-action flux spectra as a function of spacecraft-frame frequency $f_{\mathrm{sc}}$, binned by $M_a$. Panels ($\alpha$) and ($\beta$) show the spectral wave-action flux for outward- ($E_{S^{o}}$) and inward-propagating ($E_{S^{i}}$) wavepackets, respectively. Median spectra are computed within logarithmically spaced $M_a$ bins, with solid lines denoting sub-Alfv\'enic intervals ($M_a < 1$) and dotted lines corresponding to super-Alfv\'enic intervals ($M_a \geq 1$). Reference power-law slopes are indicated for comparison. Panel ($\gamma$) displays the total wave-action flux, $\mathcal{E}_{S^{\ast}}$, as a function of $M_a$ for selected frequencies. Colored points denote median values with quartile error bars; the corresponding frequencies are indicated by the colorbar, with the reference frequency marked by the white line.}\label{fig:wave_action_flux_spectra}
\end{figure*}

 % \newpage

\section{Results} \label{sec:Statistical Results}

Building on the data selection and processing described in Appendix~\ref{App:Data_Sel_and_Processing}, we first summarize the dataset's key statistical properties that motivate the additional selection criteria used in the main analysis. We then characterize the fluctuation spectra. To assess the relative contributions of linear and nonlinear processes, we analyze the wave-action flux $S^{\ast}$ alongside complementary diagnostics: the volumetric heating rate $Q$ (Appendices~\ref{Appendix_Theory:CH09} and~\ref{Appendix_Theory:Cr09}), the turbulent cascade rate $\epsilon$ (Appendix~\ref{Appendix_Theory:PP98}), and the Alfv\'en-wave energy flux $F_a$ (Appendix~\ref{Appendix_Theory: Ffuxes}).

%All quantities are computed using the five-point increment method (Appendix~\ref{Appendix_methods:5pt_Increments}), ensuring consistency across diagnostics.

\subsection{Dataset Overview and Additional Selection Criteria} \label{sec:Results_dataset_overview}

Figure~\ref{fig:heatmaps} presents bin-averaged maps of solar-wind parameters as functions of heliocentric distance and key flow variables. Panels~($\alpha$) and~($\beta$) show the variation of the mass flux, $M_f \equiv \rho U R^2$, and plasma beta, $\beta \equiv n_{p} K_{B} T / (B^{2}/2 \mu_{0})$, across the $(R,~M_a)$ plane. Panel~($\gamma$) displays the distribution of $M_a$ in the $(R,~U)$ plane. In all panels, the color scale represents the bin-averaged quantity using a logarithmic normalization, while black contours indicate the logarithmic sampling density.

The mass flux $M_f$ remains approximately constant with radial distance, with typical values of order $10^8~\mathrm{kg,s^{-1}}$. However, a distinct population of intervals with anomalously low $M_f$ and low plasma $\beta$ appears at small Alfv\'en Mach numbers ($M_a \lesssim 0.25$) and radial distances of $\sim 15 \:-\:18R_{\odot}$. As shown in panel($\gamma$), these intervals also exhibit reduced solar-wind speeds, deviating from the otherwise well-defined, monotonically increasing baseline velocity profile. Although such low-flux intervals occur sporadically across the $(M_a,~R)$ domain, they become more prevalent at low $M_a$, reflecting the shared dependence of both $M_f$ and $M_a$ on plasma density and bulk flow speed. At larger $R$ and higher $M_a$, this anomalous population is largely obscured by the dominant, steady-state wind.

Time-series inspection indicates that intervals exhibiting suppressed values of the mass-flux $M_f$, plasma beta $\beta$, and solar wind speed $U$ are predominantly associated with transient events-such as coronal mass ejections (CMEs)-which give rise to localized plasma rarefaction regions \citep[e.g.,][]{Romeo_2023, Jagarlamudi_2025}; see also Appendix~\ref{Appendix_methods:WKB_adherence}. To mitigate contamination from such events, all intervals falling below the 0.1\textsuperscript{th} percentile of the $M_f$ distribution.

Following this selection, the dataset is partitioned into $N_x = 100$ logarithmically spaced radial bins. Within each bin, the data are further divided into $N_y = 4$ equal-population quantiles, $Q_i(R)$ for $i = 1, \dots, 4$. The boundaries between quantiles are fitted with isothermal Parker-wind velocity profiles, following the procedure outlined in Appendix~\ref{Appendix:methods_quantiles}. These fits, $U_i(R)$, provide radial-dependent velocity thresholds used to classify each measurement into one of four distinct acceleration profiles. Each profile is then analyzed independently to determine the mean fluctuation properties as a function of $M_a$.

\begin{figure*}
        \centering
         \includegraphics[width=1\textwidth]{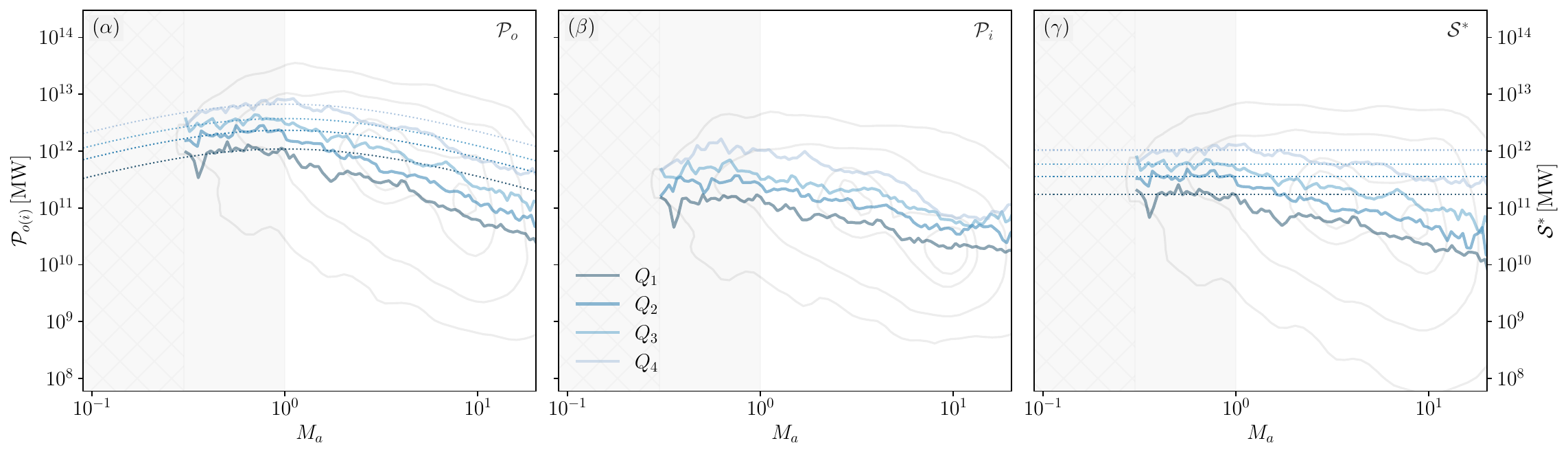}
        \caption{Power in ($\alpha$) outward-propagating, $\mathcal{P}_{o}$, and ($\beta$) inward-propagating, $\mathcal{P}_{i}$, Els\"asser fluctuations, and ($\gamma$) total wave-action flux, $\mathcal{S}^\ast$, computed at a fixed increment scale $\tilde{\ell}=1024 \ d_i$ and plotted versus $M_a$. Curves show means in logarithmic $M_a$ bins, evaluated separately for the four speed quantiles $Q_i(R)$ listed in Table~\ref{tab:parker_fit}. In ($\alpha$), dashed lines are best-fit WKB profiles $\propto M_a/(M_a+1)^2$ over $M_a<1$; in ($\gamma$), they indicate the median $\mathcal{S}^\ast$ over the same range.  Black contours denote sampling density at $[0.5\sigma,~1\sigma,~2\sigma,~3\sigma]$.
}\label{fig:Power_WKB}
\end{figure*}

\newcommand{\tblfrac}{0.9\columnwidth}

\begin{table}[htbp]
\centering
\small
\setlength{\tabcolsep}{4pt}
\begin{tabular*}{\tblfrac}{@{\extracolsep{\fill}} ccc }
\toprule
$E_i$ & $C$ (km\,s$^{-1}$) & $R_c$ ($R_{\odot}$) \\
\midrule
$E_{1}$ &  76.4175 & 7.1206   \\
$E_{2}$ &  80.1767 & 1.9388   \\
$E_{3}$ &  90.4937 & 1.2911   \\
$E_{4}$ & 116.4553 & 1.7272   \\
$E_{5}$ & 127.9708 & 0.0209   \\
\bottomrule
\end{tabular*}
\caption{Isothermal Parker model fit coefficients for each speed quantile edge, $E_i$, where
$E_i(R;C,R_c) = C\,\sqrt{-\,W\!\bigl(-\exp[-(J(R,R_c)+1)]\bigr)}$.}
\label{tab:parker_fit}
\end{table}

\subsection{Quantifying the Imbalance, Residual Energy \& Compressibility} \label{sec:Results_joint_sigmas}

To assess the degree of Alfv\'enicity in the dataset, we compute 1-minute rolling averages of the normalized cross helicity, $\sigma_c(\ell^{\ast})$, and residual energy, $\sigma_r(\ell^{\ast})$, as defined in Appendix~\ref{Appendix:methods}. Although both diagnostics exhibit scale dependence-arising in part from instrumental limitations \citep[see, e.g.,][]{Podesta_2010, sioulasSDDA}-we fix the increment scale at $\ell^{\ast} = 4096 \ d_i$. This scale statistically corresponds to the peak of $\sigma_c$ \citep[e.g.,][]{Wicks_2013_imbalanced_obse, sioulasSDDA} and is sufficiently large to remain above the high-frequency noise floor in the velocity measurements. The position of the $\sigma_c$ peak, when normalized by $d_i$, exhibits minimal variation with either $R$ or $M_a$. In the main analysis, we focus on diagnostics computed from perpendicular increments, $\delta \xi^{\perp}$, as indicated by the superscript $^{\perp}$. For completeness, analogous results based on full-vector increments, $\delta \bm{\xi}$, are presented in Appendix~\ref{Appendix_methods:sigmas_full}.

Figure~\ref{fig:sigmas} shows the joint distributions of $\sigma^{\perp}_c$ and $\sigma^{\perp}_r$, separated into four velocity-based quartiles as defined in Table~\ref{tab:parker_fit}. Each panel displays data binned into 30 logarithmically spaced intervals in $M_a$. Markers indicate the median in each bin, with symbol shape denoting the Alfv\'enic regime: circles for sub-Alfv\'enic intervals ($M_a<1$) and squares for super-Alfv\'enic intervals ($M_a\ge 1$). Vertical and horizontal error bars show the interquartile range (25th-75th percentiles). The color scale, rendered with logarithmic normalization, indicates the median $M_a$ in each bin. Gray dashed and dash-dotted curves denote contours of constant alignment angles: Els\"asser alignment, $\cos\,\theta{z^{\pm}}$, and velocity-magnetic-field alignment, $\cos\,\theta_{ub}$, respectively.

Consistent with earlier studies in the super-alfv\'enic regime \citep[e.g.,][]{DAmicis_2015, shi_alfvenic_2021}, we find that within the sub-Alfv\'enic regime ($M_a<1$) higher solar-wind speeds are associated with an enhanced degree of Alfv\'enicity. The slow and intermediate-speed populations (panels~$(\alpha)$-$(\gamma)$) evolve from a near-Sun state that is strongly imbalanced ($\sigma^{\perp}_c \approx 1$) and only weakly magnetically dominated ($\sigma^{\perp}_r \approx -0.2$), toward a quasi-balanced ($\sigma^{\perp}_c \approx 0.3$-$0.4$), strongly magnetically dominated state ($\sigma^{\perp}_r \lesssim -0.65$) at larger $M_a$. A modest increase in $\sigma^{\perp}_c$ is nonetheless observed as the Alfv\'en surface is approached ($M_a \le 1$).

In contrast, the fast wind (panel~$(\delta)$) maintains a high degree of Alfv\'enicity over most of the $M_a$ range, with $\sigma^{\perp}_c \approx 1$ and $\sigma^{\perp}_r \approx 0$. At $M_a \gtrsim 10$, however, a slight reduction in cross helicity is accompanied by a gradual transition toward magnetic dominance.

At low $M_a$, points in the $\sigma^{\perp}_r$-$\sigma^{\perp}_c$ plane cluster near the boundary of the unit circle, reflecting strong alignment both between the counterpropagating Els\"asser fields, $\bm{z}^o$ and $\bm{z}^i$, and between velocity and magnetic-field fluctuations. Among ${\sigma_c,~\sigma_r,~\theta^{ub}_{\perp},~\theta^{z}_{\pm}}$, only two are independent; their evolution therefore directly governs the alignment geometry. As $M_a$ increases, both $\sigma_c$ and $\sigma_r$ systematically decrease, consistent with increasing misalignment of the corresponding field increments. The implications of these trends are examined in Section~\ref{sec:Dicussion}.

% This pattern closely resembles predictions from the reflection-driven turbulence model of \citet{meyrand2023reflectiondriven}, where turbulence evolution is governed by anomalous coherence effects under conditions of large expansion-to-nonlinear time ratio, $\chi_{\mathrm{exp}}$. Notably, the RMS amplitudes of $\tilde{\bm{z}}^+$ fluctuations in the fast wind at $M_a \approx 7$-$8$ are comparable to those in the slow wind at $M_a \lesssim 1$, suggesting that both regimes may satisfy conditions for this phenomenology. These observations support the qualitative validity of the reflection-driven model across a broader range of solar wind streams than originally emphasized, although the fast wind shows weaker evolution, possibly reflecting the more constrained nature of wave reflection in that regime.

\begin{figure*}
    \centering
         \includegraphics[width=0.99\textwidth]{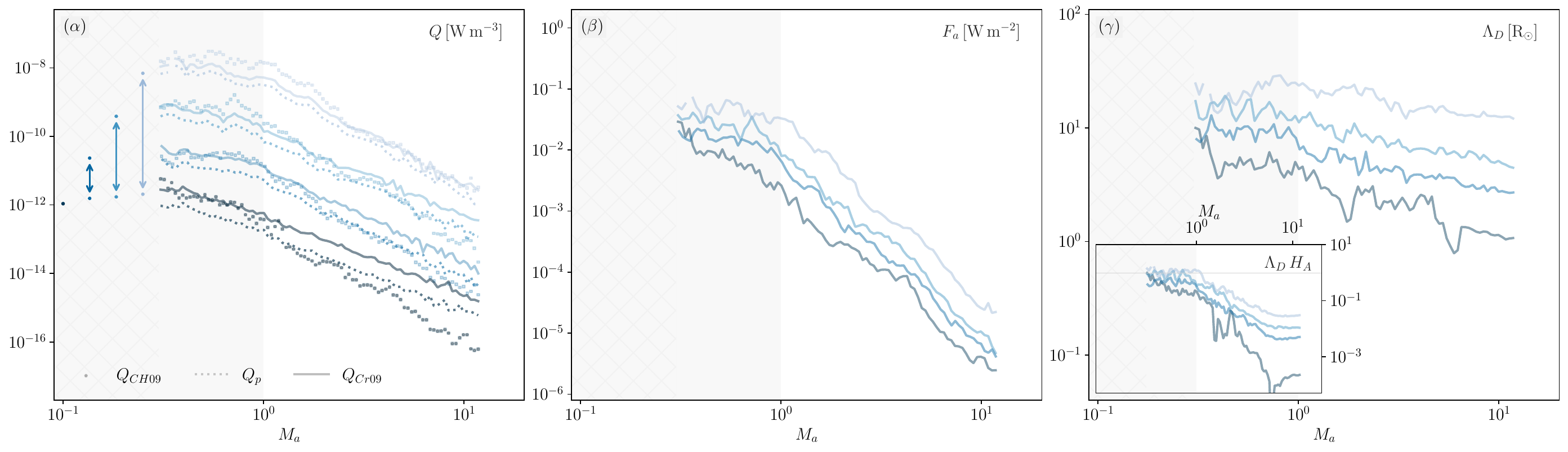}
        \caption{($\alpha$) Volumetric heating rates, $Q$, estimated using the \citetalias{Cr09} model (solid lines) and the \citetalias{CH09} model (scatter points), plotted as a function of the Alfv\'en Mach number, $M_a$. The proton-only component, $Q_p$, from the \citetalias{Cr09} model is shown with dotted lines. For clarity, profiles corresponding to different solar-wind speed bins are vertically offset by factors of $10^n$, where $n = 0$ (dark blue) denotes the slowest wind bin and $n = 3$ (light blue) the fastest; offset values are indicated by the reference bars on the side.  ($\beta$) Alfv\'en-wave energy flux.  ($\gamma$) Turbulent damping length, $\Lambda_d = F_a / Q_{\mathrm{Cr09}}$. The inset figure illustrates $\Lambda_d$ normalized by the inverse Alfv\'en-speed scale height, $\left[ \frac{d}{dR} \ln V_{a}^{1/2} \right]^{-1}$.
}\label{fig:Q_Ld_fa}
\end{figure*}

% \newpage
\subsection{Wave-Action Flux: Spectral Properties} \label{subsec:Results_wave_act_spectra}

Figures~\ref{fig:wave_action_flux_spectra}$\alpha$--$\beta$ display the wave-action-flux spectra of outward- and inward-propagating wave packets,
$E_{S^{o}}$ and $E_{S^{i}}$, defined in Equation~\eqref{eq:wave_action_flux_spectra_definition}.  
Spectra are sorted by Alfv\'en-Mach number, $M_{a}$, and plotted versus spacecraft-frame frequency, $f_{\mathrm{sc}}$.  
Because the conversion from the frequency-compensated Els\"asser spectra, $E_{o(i)}\,f_{\mathrm{sc}}$, to $E_{S^{o(i)}}$ involves only frequency-independent factors (Appendix~\ref{Appendix_methods:wave_action_flux_spectra}), the spectral slopes are preserved.

In the sub-Alfv\'enic regime, $M_{a}<1$, the outward directed spectrum, $E_{S^{o}}$, shows no significant dependence on $M_{a}$: all curves collapse onto a common profile-the total flux
$E_{S^{\ast}}=E_{S^{o}}-E_{S^{i}}$ inherits this invariance.
By contrast, $E_{S^{i}}$ decreases steadily as $M_{a}\!\to\!1$, reflecting the $(U-V_{A})$ factor in its definition.  
In this regime $E_{S^{o}}\propto f_{\mathrm{sc}}^{\alpha_{z_{o}}}$ with $\alpha_{z_{o}}\approx0.5$, while
$E_{S^{i}}\propto f_{\mathrm{sc}}^{\alpha_{z_{i}}}$ with $\alpha_{z_{i}}\approx-1$
(\,$E_{o}\propto f_{\mathrm{sc}}^{-0.5}$ and $E_{i}\propto f_{\mathrm{sc}}^{-2}$\,).
At higher frequencies the inward spectrum flattens toward $\alpha_{z_{i}}\approx0$, a trend likely driven by high-frequency velocity-noise contamination under strong imbalance \citep{PODESTA_2009_SDDA,sioulasSDDA}.

Closer to the Sun the outward and inward Els\"asser spectra possess distinct shapes across the full frequency range.  
As $M_{a}$ increases these shapes converge, and the two spectra acquire similar slopes.  
This behaviour agrees with previous in-situ reports of radial spectral evolution \citep{Grappin_1990_helios,chen_evolution_2020,shi_alfvenic_2021,sioulas_22_spectral_evolu,mcintyre2023properties} and with numerical simulations \citep{Verdini_2007,Perez_Chandran_2013,Chandran_Perez_2019,2023_Chen_compres}.  
A detailed investigation of the governing processes is beyond the scope of the present work.

\begin{table}[h!]
\centering
\small
\caption{Posterior median estimates and $68\%$ credible intervals for the normalization constant $\kappa$, the transition Alfv\'en-Mach number $M_{a}^{b}$, and the post-break power-law slope $\beta$, inferred separately for each quartile, $Q_i$.}
\label{table:bayes_power_results_no_gamma}
\setlength{\tabcolsep}{1.5pt}
\renewcommand{\arraystretch}{1.3}
\begin{tabular*}{\columnwidth}{@{\extracolsep{\fill}} l c c c}
\toprule
Quartile & $\kappa$ & $M_{a}^{b}$ & $\beta$ \\
\midrule
Q1 & $4.15{\times}10^{12}\,(^{+1.52}_{-1.41}\!{\times}10^{11})$ 
   & $0.924\,(^{+0.042}_{-0.042})$ 
   & $1.16\,(^{+0.023}_{-0.023})$ \\[2pt]

Q2 & $8.92{\times}10^{12}\,(^{+2.37}_{-2.21}\!{\times}10^{11})$ 
   & $0.926\,(^{+0.026}_{-0.025})$ 
   & $1.19\,(^{+0.015}_{-0.015})$ \\[2pt]

Q3 & $1.40{\times}10^{13}\,(^{+3.96}_{-3.84}\!{\times}10^{11})$ 
   & $1.06\,(^{+0.046}_{-0.044})$ 
   & $1.14\,(^{+0.022}_{-0.022})$ \\[2pt]

Q4 & $2.55{\times}10^{13}\,(^{+6.82}_{-6.45}\!{\times}10^{11})$ 
   & $1.80\,(^{+0.098}_{-0.092})$ 
   & $1.16\,(^{+0.027}_{-0.026})$ \\
\bottomrule
\end{tabular*}
\end{table}

\newpage
\subsection{Wave Action Flux: Five-Point Increments} 

We compute 1-minute rolling averages of the outward- and inward-directed wave power, $\mathcal{P}_{o(i)}(\tilde{\ell})$, as defined in Equation~\ref{eq:WKB_Power_Ma}, along with the wave-action flux, $\mathcal{S}^{\ast}(\tilde{\ell})$, from Equation~\ref{eq:wave_action_conv}, both evaluated at the fixed scale $\tilde{\ell} = 1024\,d_i$. This choice ensures coverage of spacecraft-frame frequencies within the inertial range across all heliocentric distances \citep[see, e.g.,][]{sioulas_turb_22_no1}. For brevity, we omit the explicit $\tilde{\ell}$ dependence and refer to these quantities as $\mathcal{P}_{o(i)}$ and $\mathcal{S}^{\ast}$.

The dataset is divided into four velocity quantiles, as defined in Section~\ref{sec:Results_dataset_overview}. Within each, we compute the mean of $\mathcal{P}_{o(i)}$ and $\mathcal{S}^{\ast}$ in 100 logarithmic bins over $M_a \in [0.3,~20]$.

Figures~\ref{fig:Power_WKB}$\alpha$-$\gamma$ show the resulting $M_a$ dependence. Dark-blue and light-blue curves denote the lowest ($n = 0$) and highest ($n = 3$) velocity quantiles, respectively.

For $M_a < 1$, $\mathcal{P}_{o}$ closely follows the WKB prediction (dashed lines), while $\mathcal{S}^{\ast}$ remains approximately constant within the $1\sigma$ spread (not shown). These results indicate that the outward-propagating component experiences weak reflection and negligible damping-implying that linear wave dynamics dominate over nonlinear dissipation-in the sub-Alfv\'enic regime.

Beyond the Alfv\'en surface ($M_a > 1$), $\mathcal{P}_{o}$ decays more steeply than  from WKB theory, while $\mathcal{S}^{\ast}$ decreases monotonically with $M_a$, exhibiting an approximate power-law scaling. In contrast, $\mathcal{P}_{i}$ declines more gradually, resulting in increasingly balanced fluxes of counterpropagating fluctuations as $M_a$ increases.

To determine the Alfv\'en-Mach number $M_{a}^{b}$ at which $\mathcal{P}_{o}$ deviates from the WKB prediction, we employ a Bayesian inference framework. In this approach, $\mathcal{P}_{o}$ is represented as a piecewise function of $M_a$, following WKB-like scaling at low $M_a$ and transitioning to a power-law decay beyond $M_{a}^{b}$. Appendix~\ref{Appendix_methods:bayesian_inference} outlines the full inference procedure, including an illustrative fit for $Q_3$. Posterior estimates, and associated credible intervals for all fitted parameters are presented in Table~\ref{table:bayes_power_results_no_gamma}.

A similar trend is observed in the radial profile of $\mathcal{P}_{o}(R)$ (not shown), which follows a single power-law scaling across the full radial range. Unlike $\mathcal{S}^{\ast}(R)$, no clear turnover-i.e., a peak followed by a monotonic decline-is observed at small $R$. This difference likely reflects the spatial complexity of the Alfv\'en surface, now understood not as a sharp boundary but as an extended and fragmented transition region \citep{deforest_highly_2018, chhiber2022MNRAS}. As a result, intervals at fixed $R$ may include both sub- and super-Alfv\'enic plasma, smoothing features in radial profiles such as $\mathcal{S}^{\ast}(R)$. An alternative interpretation is discussed in Section~\ref{subsec:dicussion_wave_action_reasons}.

\subsection{Volumetric Heating Rates, AW Fluxes and Turbulent Damping Length} \label{subsec:heating_Rates}

This section examines the dissipation of Alfv\'enic wavepackets as a function of $M_a$ and evaluates their contribution to solar wind heating. The volumetric dissipation rate $Q$ is estimated using three independent methods: (i) the reflection-driven, imbalanced MHD turbulence model of \citetalias{CH09} (Appendix~\ref{Appendix_Theory:CH09}), which defines $Q_{\mathrm{CH09}}$; (ii) the empirical heating prescription of \citetalias{Cr09} (Appendix~\ref{Appendix_Theory:Cr09}), which partitions energy between protons and electrons, $Q_{\mathrm{Cr09}} = Q_p + Q_e$; and (iii) the third-order structure-function estimate of \citetalias{PP98} (Appendix~\ref{Appendix_Theory:PP98}), interpreted as the small-scale dissipation rate, $Q_{\mathrm{PP98}}$. Figure~\ref{fig:Q_Ld_fa}$\alpha$ displays $Q_{\mathrm{CH09}}$ (dashed) and $Q_{\mathrm{Cr09}}$ (solid); $Q_{\mathrm{PP98}}$ closely follows $Q_{\mathrm{CH09}}$ (Figure~\ref{fig:heat_rates_distance}) and is omitted to reduce visual clutter.

Heating rates are computed following the methodology outlined in Appendix~\ref{Appendix_methods:heating rates}. Due to the unavailability of electron heat-flux measurements in the \citetalias{muller_solar_2020} dataset, the analysis is limited to intervals from \citetalias{fox_solar_2016}. Consequently, the sample size decreases substantially for $M_a > 12$, and these bins are excluded. Profiles corresponding to different solar-wind speed intervals are vertically offset by factors of $10^n$ ($n = 0 \dots 3$), with the offset magnitudes indicated by reference bars.

The heating rates exhibit a modest correlation with $U$. As functions of heliocentric distance, they follow a power-law decay consistent with earlier results \citep{Bandyopadhyay_2023, Bowen_2025}. By contrast, when plotted against $M_a$, the profiles $Q(M_a)$ tend to flatten for $M_a \lesssim 1$, especially in the faster streams.

Agreement among dissipation estimates depends on both $U$ and $M_a$. In the two fastest intervals ($Q_3$, $Q_4$), the three methods---$Q_{\mathrm{Cr09}}$, $Q_{\mathrm{CH09}}$, and $Q_{\mathrm{PP98}}$---agree within $1\sigma$ across the full $M_a$ range. In the slower intervals ($Q_1$, $Q_2$), the estimates coincide at low $M_a$ but diverge at higher $M_a$, with $Q_{\mathrm{Cr09}}$ consistently exceeding the incompressible estimates.

Three interpretations are plausible:

(a) The \citetalias{CH09} model assumes strong imbalance and attributes heating solely to the dominant outward-propagating component, neglecting dissipation associated with the ingoing Els\"asser field. As cross helicity decreases with increasing $M_a$---and with decreasing $U$---this assumption becomes invalid, leading the model to underestimate the true dissipation rate in more balanced intervals.

(b) Increasing compressibility \citep{ chen_evolution_2020, shi_alfvenic_2021}, particularly at larger heliocentric distances and in slower wind, may activate damping channels \citep[see e.g., ]{Barnes_1966_damping, Schekochihin_2009_review} not captured by the incompressible frameworks underlying both \citetalias{CH09} and \citetalias{PP98}, resulting in systematic underestimation by these models relative to $Q_{\mathrm{Cr09}}$.

(c) Uncertainties in particle measurements increase with heliocentric distance and may bias $Q_{\mathrm{Cr09}}$ high at large $M_a$. However, this explanation appears less likely, as the discrepancy is confined to slower wind; if instrumental effects were dominant, similar deviations would be expected in the faster intervals as well. As described in Appendix~\ref{Appendix_Data:PSP}, mitigation procedures have been applied to reduce these particle-data uncertainties.

Regardless of interpretation, the discrepancy largely disappears when considering only the proton contribution, $Q_p$, suggesting that a growing fraction of turbulent energy is transferred to electrons as the turbulence becomes more balanced and the plasma $\beta$ increases. This trend is consistent with recent observational evidence \citep{Shi_2023_ion_elec} and with theoretical predictions linked to the ``helicity-barrier'' mechanism \citep{Meyrand_transition, 2022_Squire_helicity, Squire_2023}.

Figure~\ref{fig:heat_rates_distance}b shows the total Alfv\'en-wave energy flux, $F_A$, which follows a broken power-law profile with a shallower decay in the sub-Alfv\'enic regime. Normalizing by the dissipation rate $Q_{\mathrm{Cr09}}$ defines a turbulent damping length, $\Lambda_D = F_A / Q_{\mathrm{Cr09}}$, interpreted as the characteristic distance over which AW energy is dissipated.

Figure~\ref{fig:Q_Ld_fa}$\gamma$ shows that $\Lambda_D$ decreases with increasing $M_a$ within each wind-speed bin and remains consistently larger in faster streams. The decline is modest for $M_a < 1$ but becomes significantly steeper beyond the Alfv\'en point, suggesting enhanced dissipation at larger heliocentric distances.

To assess the role of reflection-driven turbulence, we compare $\Lambda_D$ to the reflection scale $H_A \equiv \left(d\ln v_A^{1/2}/dR\right)^{-1}$, set by the background Alfv\'en-speed gradient. If reflection dominates the cascade, one expects $\Lambda_D \sim H_A$. The inset of Figure~\ref{fig:Q_Ld_fa}$\gamma$ shows the dimensionless ratio $\Lambda_D / H_A$. Values are near unity in fast streams and for $M_a \lesssim 1$, but fall below unity as $M_a$ increases and in slower wind. This departure implies that reflection alone cannot account for the observed damping lengths across much of the parameter space. The systematic decline of $\Lambda_D / H_A$ with increasing $M_a$ and decreasing $U$ suggests that additional dissipation mechanisms---such as compressible damping---not accounted for by incompressible RDT become increasingly important.

\section{Discussion} \label{sec:Dicussion}

This section places the present results in context with established theory, prior spacecraft observations, and numerical simulations. Points of agreement and divergence are identified to delineate the implications for AW evolution and the role of Alfv\'enic turbulence in governing solar-wind dynamics.

\subsection{Conservation of Wave Action}\label{subsec:dicussion_wave_action_reasons}

The analysis indicates that the wave-action flux, $S^{\ast}$, remains approximately conserved ($S^{\ast} \approx \text{const.}$) up to a characteristic Alfv\'en Mach number, $M_{a}^{b}$. In this regime, the outward-directed power, $\mathcal{P}_{o}$, adheres to the WKB-predicted scaling. For $M_{a} > M_{a}^{b}$, both $S^{\ast}$ and $\mathcal{P}_{o}$ transition to a steeper power-law decay. While $M_{a}^{b}$ typically lies near the critical Alfv\'en surface ($M_{a}^{b} \approx 1$), its value varies systematically with solar wind speed: in the slowest streams, the break occurs below the Alfv\'en surface ($M_{a} < 1$), whereas in faster flows it shifts to $M_{a} \gtrsim 1$ (see Table~\ref{table:bayes_power_results_no_gamma}).

The observed trend reflects the established positive correlation between normalized cross helicity, $\sigma_c$, and bulk solar wind speed, $U$ (Figure~\ref{fig:sigmas}). In slow wind, low $\sigma_c$ indicates near balance between counter-propagating Alfv\'enic fluxes, which strengthens nonlinear interactions and drives earlier deviations from WKB behavior \citep{Verdini_2007, Verdini_2010, Chandran_Perez_2019}. In contrast, fast streams are characterized by strong outward dominance, leading to elevated $\sigma_c$ and close alignment between velocity and magnetic field fluctuations \cite[see also][]{Perez_2009, sioulasSDDA}. This alignment suppresses nonlinear coupling \citep{boldyrev_2006, Perez_Boldyrev_2009, chandran_intermittency_2015, Mallet_2017}, thereby permitting outward-propagating fluctuations to retain WKB-like scaling over a broader range in $M_a$.

Conditional sub-sampling supports this interpretation: when the analysis is restricted to strongly imbalanced intervals ($\sigma_c \geq 0.75$), the scaling of $\mathcal{P}_o$ with $M_a$ remains consistent with WKB predictions across the full range probed (not shown).

It should be noted, however, that $M_{a}^{b}$ exhibits a clear dependence on fluctuation scale, as shown in Figure~\ref{fig:wave_action_flux_spectra}$\gamma$. The reported $M_{a}^{b}$ values are therefore scale-specific, and this dependence must be accounted for in any future cross-dataset comparison or interpretation.

Two additional factors merit attention, as they help explain the more gradual decline of $S^{\ast}$ with $M_a$ in the sub-Alfv\'enic regime, in contrast to the steeper decay observed beyond the Alfv\'en surface.

First, below the Alfv\'en point, $M_a$ increases rapidly with radial distance due to the steep drop in plasma density during the solar wind acceleration phase. For a radial magnetic field, $M_a \propto n^{-1/2}$, and since the density scales as $n \propto R^{-a}$ with $a > 2$, this implies a rapid increase in $M_a$ with $R$ for $M_a < 1$ \citep{Chandran_Perez_2019}. Consequently, even if the wave-action flux $S^{\ast}$ varies similarly with $R$ on either side of the Alfv\'en surface, its dependence on $M_a$ appears shallower in the sub-Alfv\'enic range due to the nonlinear mapping between $R$ and $M_a$.

Second, in reflection-driven Alfv\'en wave turbulence, the dissipation rate scales inversely with the Alfv\'en speed scale height, $H_A \sim |V_A / (dV_A/dr)|$. In the sub-Alfv\'enic region, $H_A$ tends to be large relative to $R$, implying reduced dissipation efficiency. This is because, although the magnetic field decreases as $B \propto R^{-2}$, the plasma density falls off even more steeply, resulting in a weak radial gradient of the Alfv\'en speed: typically $V_A \propto R^{-n}$ with $n < 1$. The shallow $V_A$ profile produces a large $H_A$, thereby limiting the rate at which turbulent energy is dissipated below the Alfv\'en surface.

While the preceding caveats impose important constraints on interpretation, one result is unaffected: the present findings differ qualitatively from those of \citet{Ruffolo_2024}, who identified a localized enhancement of wave-action flux near the Alfv\'en surface. This discrepancy reflects several methodological differences, each of which significantly alters the inferred fluctuation profiles.

A primary methodological distinction lies in the treatment of stream heterogeneity. \citet{Ruffolo_2024} analyze a single, aggregated dataset spanning a wide range of wind speeds and dynamical conditions. In contrast, the present analysis stratifies the data either by acceleration profile or into narrow bins of approximately constant bulk velocity (Appendix~\ref{appendix_furth_res_U_bins}). This classification reduces cross-contamination between streams with distinct thermodynamic and dynamical profiles---an essential constraint, given the established dependence of turbulence amplitudes \citep{Vasquez_2007, Pi_2020} and their radial evolution \citep{shi_alfvenic_2021, Dakeyo_2022, Adhikari_2022, sioulas_22_spectral_evolu} on solar wind speed. 

A second source of divergence arises from the definition of fluctuation amplitudes. \citet{Ruffolo_2024} compute fluctuations relative to a 1-minute rolling average, thereby imposing a high-pass filter that suppresses low-frequency variability and removes a substantial fraction of the total power. Since wave-action flux is dominated by large-scale fluctuations, this filtering introduces a systematic bias that directly affects the inferred profile.

Third, the present study excludes intervals characterized by anomalously low mass flux, which are typically associated with transient events such as coronal mass ejections or heliospheric current sheet crossings \citep{Romeo_2023, Jagarlamudi_2025}. These events violate the assumptions of steady-state flow and, if included, can significantly distort estimates of fluctuation amplitudes near the Alfv\'en surface. 

Finally, the temporal and spatial coverage differs substantially. The present dataset spans Encounters E1 through E24 \citepalias{fox_solar_2016}, while \citet{Ruffolo_2024} restrict their analysis to E8-E17, a phase corresponding primarily to the rising portion of the solar cycle. As the geometry of the Alfv\'en surface evolves with solar activity \citep{finley_Alfven_surface, badman_2025_alfven_surface}, this difference may affect the comparison. However, restricting the present analysis to the same E8-E17 interval yields consistent results, provided the classification and filtering procedures described above are maintained.

\subsection{On the Origin of Negative Residual Energy: Turbulent, Linear, and Instrumental Contributions}

The analysis consistently yields negative values of $\sigma_r$, indicating a modest excess of magnetic over kinetic fluctuation energy \citep[see also][]{chen_evolution_2020, shi_alfvenic_2021}. In ideal, homogeneous MHD, Alfv\'en waves exhibit exact equipartition between velocity and magnetic fluctuations, corresponding to $\sigma_r = 0$ \citep{Walen_1944}. However, a growing body of theoretical and numerical work demonstrates that nonlinear interactions between counter-propagating Alfv\'enic wavepackets generically produce a persistent magnetic excess as energy cascades to smaller scales \citep{Muller_grappin_2005, boldyrev2012residual, howes2013alfven, dorfman2024residual}.

As noted by \citet{Ruffolo_2024}, the departure from equipartition---particularly when considered alongside the elevated heating rates reported in Section~\ref{subsec:heating_Rates}---implies a breakdown of the assumptions underlying linear wave theory. This deviation directly challenges the applicability of the WKB approximation in this regime.

Although negative residual energy is often associated with nonlinear Alfv\'enic turbulence, certain linear processes can also contribute. Solar wind expansion induces differential decay between magnetic and velocity fluctuations, leading to $\sigma_r < 0$ even in the absence of strong nonlinear coupling \citep{Dong_2014, 2023_Chen_compres}. In addition, a positive Alfv\'en speed gradient ($dV_A/dR > 0$)---as present over the radial intervals examined here---can drive non-WKB reflection that preferentially amplifies magnetic fluctuations, thereby generating negative residual energy \citep{Chandran_Perez_2019}.

In addition, measurement uncertainties in key plasma parameters can influence the observed residual energy. Proton densities derived from quasi-thermal noise spectroscopy carry uncertainties of up to $\sim 8\%$ (D. Larson, pers. comm.), and SPAN-I may under-sample the proton velocity distribution function. Although mitigation procedures are applied (Appendix~\ref{Appendix_Data:PSP}), these limitations can affect the accuracy of $\sigma_r$, including at small heliocentric distances. Such instrumental effects may partially account for the modestly negative values observed across the sampled intervals.

These considerations complicate the interpretation of residual energy as a definitive signature of nonlinear turbulent dynamics. Although the observed departure from equipartition is consistent with a magnetic-energy excess generated by a turbulent cascade, linear processes and measurement uncertainties provide alternative explanations that cannot be excluded. Accordingly, whether such departures constitute a genuine breakdown of WKB-based models remains an open question.

\subsection{On the Evolution of the Joint Distribution of Normalized Cross-Helicity and Residual Energy}

A direct comparison between our in-situ observations and the 
expanding-box \citep{Velli_1992_EBM, Grappin_velli_mengeney_1993} reduced MHD \citep[RMHD;][]{Kadomtsev_Pogutse} numerical simulations and reflection-driven turbulence model of \citet{meyrand2023reflectiondriven} is instructive for interpreting the evolution of $(\sigma_c, \sigma_r)$. In \citet{meyrand2023reflectiondriven} the key control parameter is the ratio of expansion to nonlinear timescales,
$\chi_{\exp} \equiv \tau_{\exp} / \tau_{\mathrm{nl}} \sim (a / \dot{a}) \cdot (z^{o}_{\mathrm{rms}} / \lambda_o).$  
While the outward-propagating field $\tilde{\bm{z}}^{o}$ initially dominates, its minority counterpart $\tilde{\bm{z}}^{i}$ is continually regenerated via reflection, coherently and anti-aligned with $\tilde{\bm{z}}^{o}$,  
$\tilde{\bm{z}}^{i} \sim -\tilde{\bm{z}}^{o} / \chi_{\exp}$, constraining the evolution to remain near the edge of the $\sigma_r$-$\sigma_c$ circle. Here, tildes denote wave-action-normalized fields in the expanding-box frame. Within this framework Alfv\'enic collisions are suppressed, but the “anomalous coherence” \citep{velli_turbulent_1989} permits a strong cascade even under strong imbalance \citep{Lithwick_2007_imbalanced_critical_balance, Chandran_2025}. For $\chi_{\exp} \gg 1$, the imbalance remains maximal. As $\chi_{\exp} \to 1$, the model predicts a rapid drop in $\sigma_c$ and increasingly negative $\sigma_r$, marking the onset of magnetically dominated states and reduced dissipation.

The observational results show a clear contrast between slow/intermediate and fast solar wind. In the lower-speed quartiles ($Q_1$-$Q_3$), the distribution of $(\sigma_c^\perp, \sigma_r^\perp)$ broadens and shifts away from the Alfv\'enic limit near $(1, 0)$ toward the lower part of the circle, reaching $\sigma_r^\perp \sim -0.7$. In contrast, in the fast-wind quartile ($Q_4$), the points remain tightly clustered near the right-hand edge of the circle, with $\sigma_c^\perp \simeq 1$ persisting up to $M_a \approx 10$. Beyond this point, $\sigma_c$ begins to decrease and $\sigma_r$ develops a magnetic excess.

This contrast is consistent with the amplitude trends shown in Figure~\ref{fig:appendix_RMS}: the r.m.s.\ amplitude of the outward field in $Q_4$ at $M_a \approx 10$ is comparable to that in $Q_1$ at $M_a \approx 1$, indicating that $\chi_{\exp} \gg 1$ persists to much larger $M_a$ in the fast wind. In this sense, the outward-amplitude panels act as a proxy for proximity to the reflection threshold: as long as $z^{o}_{\mathrm{rms}}/\lambda_o \gtrsim \dot a/a$ (i.e., $\chi_{\exp}\gg 1$), high imbalance persists, with the fast wind crossing this threshold at larger $M_a$. The observed behavior qualitatively matches the model prediction that imbalance persists until the outward amplitude crosses a reflection threshold, with the transition occurring at higher $M_a$ in faster wind.

When full increments are used---including the parallel components (Fig.~\ref{fig:sigmas_full_comp})---the joint distribution retracts from the unit-circle boundary at large $M_a$. This deviation arises because RMHD assumes incompressibility and transverse polarization, neglecting field-parallel structure and compressive fluctuations. The parallel content in full increments therefore mixes compressive fluctuations with field-parallel structure of large-amplitude Alfv\'enic fluctuations, none of which are represented in RMHD. Close to the Sun, where fluctuations are predominantly transverse, the choice between $\delta \xi_{\perp}$ and full $\delta\bm{\xi}$ has little effect on $(\sigma_c, \sigma_r)$. At larger $M_a$, however, increasing parallel structure makes this distinction significant: restricting to $\delta \xi_{\perp}$ biases points outward in the circle plot, while full increments shift them inward. This clarifies why perpendicular-only diagnostics more closely track RMHD predictions and remain broadly consistent with earlier analyses \citep[see e.g.,][]{sioulas_22_spectral_evolu}

 Taken together, the observations are consistent with the mechanism proposed by \citet{meyrand2023reflectiondriven}: imbalance persists until the outward-field amplitude falls below the reflection threshold, after which $\sigma_c$ declines and $\sigma_r$ becomes more negative. However, the divergence between perpendicular-only and full-increment diagnostics at large $M_a$ indicates that quantitative accuracy in this regime requires a theoretical framework that goes beyond RMHD---one that accounts for compressibility, field-parallel structure, and the full vector geometry of fluctuations in the expanding solar wind.

\subsection{Implications for sub-Alfv\'enic Switchbacks}

We conclude with brief remarks on the implications of these findings for proposed mechanisms of switchback (SB) formation. \citet{Ruffolo_2024}, citing the apparent scarcity of SBs in the sub-Alfv\'enic solar wind \citep[see also][]{Bandyopadhyay_alfven, Jagarlamudi_2023}, argued that this observation---when considered alongside a localized enhancement in wave-action flux---poses a challenge to WKB-based models of SB generation \citep[e.g.,][]{Squire_2020, Mallet_2021, squire2022b, Johnston2022}. As an alternative, they endorsed the scenario proposed by \citet{Ruffolo2020}, in which SBs are generated by energy injection from large-scale coronal shear flows near the Alfv\'en critical surface.

\par Our results support a different interpretation. The large damping lengths $\Lambda_d$ inferred in the sub-Alfv\'enic regime imply that outward-propagating Alfv\'enic fluctuations suffer relatively weak dissipation and can propagate substantial distances without significant energy loss. This is consistent with the conditions required for expansion-driven amplification, as envisioned in WKB-based SB formation models \citep{Mallet_2021, Squire2022, Squire_Mallet_2022, Johnston2022}. In forthcoming work, we will show that switchbacks do occur within sub-Alfv\'enic flows, and that this regime is particularly conducive to their formation: the deflection angle increases more rapidly below the Alfv\'en surface than above it, enhancing the likelihood of large-amplitude rotational structures.

\par Further insights will come from a scale-resolved analysis of the nonlinearity parameter, $\chi_{\lambda,\pm} \equiv \tau_{A,\pm} / \tau_{nl,\pm}$, which quantifies the ratio of linear to nonlinear timescales at each spatial scale. This will allow us to assess the strength and scale-dependence of nonlinear interactions within the sub-Alfv\'enic regime---a key determinant of switchback evolution---and will be the focus of future work.

\section{Summary and Conclusions}\label{sec:Conclusions}

We have investigated the statistical properties of large-amplitude Alfv\'enic fluctuations in the near-Sun solar wind, employing a merged dataset from \citetalias{fox_solar_2016} and \citetalias{muller_solar_2020} to span a broad range of heliocentric distances and wind regimes. Measurements were organised according to distinct solar-wind acceleration profiles, enabling a comparative analysis of outward-directed wave power, the conservation of wave-action flux, and the radial distribution of turbulent heating rates.

\vspace{1em}

The main findings can be summarised as follows:

\vspace{1em}
(1) WKB-like behaviour persists only up to a scale-dependent \emph{break Mach number}, $M_A^{b}$, beyond which the outward-propagating wave power and wave-action flux depart from WKB predictions and transition to a steeper, non-WKB power-law decay. While $M_A^{b}$ often lies near the Alfv\'en critical point ($M_A = 1$), it varies systematically with the normalised cross helicity $\sigma_c$. In slow solar wind, where $\sigma_c$ is typically low, enhanced coupling between counter-propagating Els\"asser modes leads to earlier breakdown of WKB scaling ($M_A^{b} < 1$). In contrast, in fast streams, where $\sigma_c \approx 1$, nonlinear interactions are suppressed, allowing WKB-like evolution to persist to $M_A^{b} > 1$. These trends identify $\sigma_c$ as the primary control parameter governing the extent of the WKB regime in the near-Sun solar wind.

\vspace{1em}
(2) The volumetric turbulent heating rate $Q$ peaks near the Alfv\'en critical radius $R_A$ and remains elevated primarily over $M_A \lesssim M_A^{b}$. Beyond the break ($M_A > M_A^{b}$), $Q$ decreases rapidly, indicating a preferential heating zone largely confined to regions where WKB scaling still holds \citep[see e.g.,][]{Kasper_2017}.

\vspace{1em}
(3) Despite this enhanced heating, the Alfv\'en-wave energy flux $F_a$ remains sufficiently large to yield extended damping lengths, $\Lambda_d = F_a/Q$. This implies that outward-propagating wavepackets experience relatively weak dissipation and can traverse long distances before incurring substantial energy losses, at least up to the scale-dependent threshold $M_A^{b}$.

\vspace{1em}
(4) The normalized damping length $\Lambda_d/H_A$, where $H_A$ is the inverse Alfv\'en-speed scale height, remains near unity for $M_A \lesssim M_A^{b}$ but declines systematically with increasing $M_A$. This trend is most pronounced in slow, quasi-balanced streams and indicates that incompressible reflection-driven turbulence alone cannot account for the observed dissipation. Additional damping mechanisms---such as compressible effects---are likely required.

\vspace{1em}

Collectively, these results provide quantitative constraints on the spatial and parametric limits of wave-action conservation and WKB evolution in the expanding solar wind. They show that the onset of turbulent dissipation and the breakdown of wave-like evolution occur near a transition point $M_A^{b}$, whose location is regulated by the 
by the interplay of plasma conditions, fluctuation scale, and nonlinear coupling strength---with cross helicity $\sigma_c$ emerging as a key regulator. These findings have direct implications for models of coronal heating, reflection-driven turbulence, and the generation of magnetic switchbacks in the inner heliosphere.

%\begin{acknowledgments}

\vspace{3em}
NS gratefully acknowledges insightful discussions with Jono Squire, which substantially contributed to improving the manuscript.

This research was funded in part by the FIELDS experiment
on the Parker Solar Probe spacecraft, designed and developed under NASA contract
NNN06AA01C; the NASA Parker Solar Probe Observatory Scientist
grant NNX15AF34G and the  HERMES DRIVE NASA Science Center grant No. 80NSSC20K0604. 
The instruments of PSP were designed and developed under NASA contract NNN06AA01C. 
CS was supported by NASA ECIP 80NSSC23K1064. M.L. acknowledges support from NASA HGIO grant 80NSSC25K7689.

\software{The authors acknowledge the following open-source packages: \citetalias{van1995python}, \citetalias{2020SciPy-NMeth}, \citetalias{mckinney2010data}, \citetalias{Hunter2007Matplotlib}, \citetalias{PYMC_2015}, \citetalias{angelopoulos_space_2019}, \citetalias{MHDTurbPy_Sioulas}.}
%\end{acknowledgments}

\clearpage
\appendix

\section{Theoretical Background} \label{Appendix:background}
\vspace{0.5em}

\subsection{Conservation of Wave Action}\label{Appendix_Theory:Wave_action}

For completeness, we summarise the conservation of wave action, i.e., the conservation of Alfv\'enic wave energy in the solar-wind comoving frame, following the formalism of \citet{Heinemann_Olbert}. We define the Heinemann-Olbert (HO) variables
\begin{equation}\label{eq:HO_variables}
    \boldsymbol{g}_{\pm} \equiv \frac{\boldsymbol{z}_{\pm}\,(1 \pm \eta^{1/2})}{\eta^{1/4}},
\end{equation}
where $\boldsymbol{z}_{\pm}$ are the Els\"asser fields, $U$ is the bulk flow speed, $V_A$ is the Alfv\'en speed, and $\eta \equiv (V_A/U)^2$ is the squared inverse Alfv\'en Mach number. The HO variables are constructed so as to remove the leading-order effects of the radial inhomogeneity in the background flow.

Starting from the Els\"asser formulation of the incompressible MHD equations Eq.~\eqref{eq:MHD} and setting the nonlinear term $\mathcal{F}_{\!nl}=0$, we separate the resulting transport equations into real and imaginary parts. Multiplying each equation by the corresponding Els\"asser amplitude and summing yields the conservation law for wave action:
\begin{equation}\label{eq:wave_action_conv}
    \frac{\partial}{\partial t}
    \left[ \rho U\left( \frac{g_{+}^{2}}{U+V_A} - \frac{g_{-}^{2}}{U-V_A} \right) \right]
    + \nabla \cdot \left[ \rho \,\boldsymbol{U}\,(g_{+}^{2} - g_{-}^{2}) \right] = 0,
\end{equation}
where $\rho = m n$ is the mass density.

We identify the \emph{wave-action density} as
\begin{equation}
    \mathcal{S} \equiv \rho U\left( \frac{g_{+}^{2}}{U+V_A} - \frac{g_{-}^{2}}{U-V_A} \right).
\end{equation}
Equation~\eqref{eq:wave_action_conv} can then be written as,
\begin{equation}
    \frac{\partial \mathcal{S}}{\partial t}
    + \nabla \cdot \left( \boldsymbol{c}\,\mathcal{S} \right) = 0,
\end{equation}

where $\boldsymbol{c}$ is the group velocity given by
\begin{equation}
    \boldsymbol{c} =
    \left[
        \frac{g_{+}^{2} - g_{-}^{2}}
             {g_{+}^{2}/(U+V_A) - g_{-}^{2}/(U-V_A)}
    \right] \hat{\boldsymbol{b}}.
\end{equation}
Here $\hat{\boldsymbol{b}}$ is the unit vector along the mean magnetic field.

Assuming harmonic time dependence $z_{\pm} \propto \exp(-i\omega t)$, we define the wave-action fluxes associated with antisunward ($+$) and sunward ($-$) propagation as
\begin{equation}\label{eq:S_p_m_definition}
    S^{\ast}_{\pm} \equiv M_f\,g_{\pm}^{2},
\end{equation}
where
\begin{equation}
    M_f \equiv \rho\,U\,R^2
\end{equation}
is the total radial mass flux and $R$ is the heliocentric distance. In this notation, the conservation law~\eqref{eq:wave_action_conv} becomes
\begin{equation}\label{eq:Sstar_const}
    S_{+}^{\ast} - S_{-}^{\ast} \;=\; \mathcal{S}^{\ast} \;=\; \mathrm{const}.
\end{equation}

In the limit of nearly pure antisunward propagation, $|z_{+}|^{2} \gg |z_{-}|^{2}$, the minority component $g_{-} \to 0$ and $S_{+}^{\ast}$ is itself conserved, implying $\partial_{r}g_{+}^{2} = 0$. Using Eq.~\eqref{eq:S_p_m_definition}, one obtains
\begin{equation}\label{eq:WKB_Power_Ma}
    \mathcal{P}_{o} \;\equiv\; \frac{M_f\,|z_{+}|^{2}}{4}
    \;\propto\; \frac{M_A}{(M_A+1)^{2}},
\end{equation}
where $M_A \equiv U/V_A$ is the Alfv\'en Mach number. Equation~\eqref{eq:WKB_Power_Ma} reproduces the WKB prediction for non-reflecting, antisunward-propagating Alfv\'en waves \citep{Jacques_1977}, with a global maximum in $\mathcal{P}_{o}$ occurring at the Alfv\'en critical surface ($M_A = 1$).

\subsection{Empirical Proton \& Electron Heating Rates}\label{Appendix_Theory:Cr09}

Starting from the steady-state internal-energy conservation equations for protons and electrons, one may write the volumetric heating rates as \citet[][hereafter \citetalias{Cr09}]{Cr09},

\begin{align}
Q_{p}(r) &=
\frac{3}{2}\,n_{p}U k_{B}\frac{dT_{p}}{dr}
-U k_{B}T_{p}\frac{dn_{p}}{dr}
+\frac{3}{2}\,n_{p}k_{B}\nu_{pe}(T_{e}-T_{p}), \\[6pt]
Q_{e}(r) &=
\frac{3}{2}\,n_{e}U k_{B}\frac{dT_{e}}{dr}
-U k_{B}T_{e}\frac{dn_{e}}{dr}
-\frac{3}{2}\,n_{e}k_{B}\nu_{ep}(T_{e}-T_{p})
+\frac{1}{r^{2}}\frac{d}{dr}\Bigl[r^{2}q_{\parallel,e}\cos^{2}\!\Phi(r)\Bigr],
\end{align}
where \(k_{B}\) is Boltzmann’s constant, and \(\nu_{pe}\) and \(\nu_{ep}\) are the Coulomb collision frequencies between protons and electrons:
\[
\nu_{pe}=
8.4\times10^{-9}
\left(\frac{n_{e}}{2.5\,\mathrm{cm^{-3}}}\right)
\left(\frac{T_{e}}{10^{5}\,\mathrm{K}}\right)^{-3/2}\,\mathrm{s^{-1}},
\qquad
\nu_{ep}=
8.4\times10^{-9}
\left(\frac{n_{p}}{2.5\,\mathrm{cm^{-3}}}\right)
\left(\frac{T_{p}}{10^{5}\,\mathrm{K}}\right)^{-3/2}\,\mathrm{s^{-1}}.
\]
and $\Phi(r)$ Parker spiral angle, expressed as:

\begin{equation}
\tan\Phi(r)  = \left(\frac{\omega (r - r_0)}{U_r}\right),
\end{equation}

where $\omega = 2.9 \times 10^{-6} \text{ s}^{-1}$ represents the Sun's equatorial angular velocity, $r(t)$ is the radial distance from the spacecraft to the center of the Sun, $r_0 = 10 R_{\odot}$ \citep{Bruno_1997} is the reference distance for the Parker spiral, and $U_r(t)$ denotes the observed radial velocity of the solar wind.

Finally, the total heating rate as estimated by the \citetalias{Cr09} method, is given by the sum of the proton and electron heating rates:

\begin{equation*}
    Q_{Cr09} =  Q_{p} + Q_{e}
    \label{eq:heating_rate_Cr09}
\end{equation*}

In the \citetalias{Cr09} model, temperature anisotropy effects and the proton heat conduction term are neglected, as they contribute negligibly to the overall energy balance at the distances considered. For further details and justification of these approximations, we refer the reader to \citetalias{Cr09}.

\subsection{Heating Rate in Reflection-Driven Alfv\'en Wave Turbulence \citep{CH09}}\label{Appendix_Theory:CH09}

\citet{Dmitruk_2002} developed a phenomenological model for reflection-driven turbulence in coronal holes, aimed at describing the associated turbulent heating rate. The model assumes a dominant outward wave energy flux and a sunward energy cascade timescale that is comparable to or shorter than the linear wave period.

Building on this framework, \citet[][hereafter \citetalias{CH09}]{CH09} generalised the model by accounting for the solar-wind outflow velocity, explicitly incorporating the work done by AWs on the plasma. In both formulations, the generation of sunward-propagating waves by reflection is balanced by their nonlinear dissipation through interactions with anti-sunward waves.

Starting from Equation~\eqref{eq:MHD} and adopting the key assumption
\begin{equation}
    z^+_{\text{rms}} \gg z^-_{\text{rms}},
    \label{eq:dominance}
\end{equation}
the total energy density simplifies to
\begin{equation}
    \mathcal{E}^+ + \mathcal{E}^- \simeq \frac{1}{4} \rho \left( z^+_{\text{rms}} \right)^2,
    \label{eq:energy_balance}
\end{equation}
reflecting the dominance of outward-propagating fluctuations.

Under this approximation, the turbulent heating rate can be estimated as
\begin{equation}
    Q_{CH09} = \frac{\rho \, z^-_{\text{rms}} \left( z^+_{\text{rms}} \right)^2}{4 \lambda_{\perp}},
    \label{eq:heating_rate}
\end{equation}
where $\lambda_{\perp}$ is the perpendicular correlation length of the turbulence relative to the background magnetic field $\boldsymbol{B}_0$.

To close the model, \citetalias{CH09} determined $z^-_{\text{rms}}$ by balancing its generation via non-WKB reflection with its nonlinear dissipation. In the limit of small $\lambda_{\perp}$, this yields
\begin{equation}
    z^-_{\text{rms}}(r) = \lambda_{\perp} \frac{(U + V_A)}{V_A} \left| \frac{dV_A}{dr} \right|,
    \label{eq:z_minus}
\end{equation}
implying that $z^-_{\text{rms}}$ depends only on the large-scale profiles of $U$ and $V_A$, and remains independent of $z^+_{\text{rms}}$, since both the reflection and dissipation terms scale linearly with $z^+_{\text{rms}}$.

Combining Equations~\eqref{eq:heating_rate} and \eqref{eq:z_minus}, and expressing the result in terms of the HO variables defined in Equation~\eqref{eq:HO_variables}, we obtain the \citetalias{CH09} heating rate as a decay law for large-scale Alfv\'enic fluctuations:
\begin{equation}
    Q_{\text{CH09}} = -\frac{\rho}{4(1 + \eta^{1/2})} g_{+}^{2} \frac{dV_A}{dr}.
    \label{eq:heating_rate_CH09}
\end{equation}

\subsection{Cascade-rate estimate based on the \cite{PP98} law}\label{Appendix_Theory:PP98}

For incompressible, homogeneous, and statistically steady MHD turbulence, \citet[][hereafter \citetalias{PP98}]{PP98} generalized the Kolmogorov-Yaglom third-order law to the Els\"asser variables.
Under the additional assumptions that (i) large-scale shear is negligible,  
(ii) the fluctuations are axisymmetric about the local mean field, and  
(iii) the separation scale $l$ lies in the inertial range,  
the exact relation reads
\begin{equation}
Y_{\pm}(l) \;=\; 
-\frac{4}{3}\,\epsilon^{\pm}\,l, 
\qquad
Y_{\pm}(l)\equiv
\bigl\langle \hat{\boldsymbol{l}}\!\cdot\!\delta\boldsymbol{z}_{\mp}\,
|\delta\boldsymbol{z}_{\pm}|^{2}\bigr\rangle ,
\label{eq:PP98_exact}
\end{equation}
where, $\hat{\boldsymbol{l}}=\boldsymbol{l}/l$, and
$\epsilon_{\pm}$ are the cascade (energy-transfer) rates of the $\boldsymbol{z}_{\pm}$ fields. 

The total turbulent energy cascade rate is then
\begin{equation}
Q_{\text{PP98}} \;=\;
\frac{\epsilon^{+}+\epsilon^{-}}{2},
\label{eq:PP98_Q}
\end{equation}
which we equate to the volumetric heating rate removed at small scales.

\subsection{Estimating AW Energy Fluxes}\label{Appendix_Theory: Ffuxes}

For outward-propagating Alfv\'en waves obeying wave-action conservation, the wave energy flux is given by  
\begin{equation}
F_{A} = \mathcal{E}_{+} \left( \frac{3}{2}\,U + V_A \right),
\end{equation}
where $\mathcal{E}_{+} = \tfrac{1}{4} n_p m_p \langle \delta z_+^{2} \rangle$ denotes the energy density of outward Els\"asser fluctuations. 

% The bulk kinetic energy flux is defined as  
% \begin{equation}
% F_k = \tfrac{1}{2} n_p m_p U^3,
% \end{equation}
% while the proton thermal energy fluxes associated with the perpendicular and parallel temperature components are given by  
% \begin{equation}
% F_{T_\perp} = n_p U T_{\perp p}, \quad 
% F_{T_\parallel} = \tfrac{3}{2} n_p U T_{\parallel p},
% \end{equation}
% where $T_{\perp p}$ and $T_{\parallel p}$ represent the proton temperatures perpendicular and parallel to the magnetic field, respectively. The total kinetic and enthalpy energy flux of the protons is thus  
% \begin{equation}
% F_U = F_k + F_{T_\perp} + F_{T_\parallel}.
%\end{equation}

\section{Data Selection \& Processing} \label{App:Data_Sel_and_Processing}
\vspace{0.5em}

\subsection{Parker Solar Probe (\citetalias{fox_solar_2016})} \label{Appendix_Data:PSP}

Observations from the \citetalias{fox_solar_2016} mission are analyzed over the interval 1~October~2018 to 23~June~2025, spanning the first 24 perihelion encounters (E1--E24). Magnetic-field measurements are taken from the Level~2 Fluxgate Magnetometer (FGM) data products \citep{bale_fields_2016} in RTN coordinates. Proton plasma moments are derived from Level~3 Solar Wind Electrons Alphas and Protons (SWEAP) data. For heliocentric distances $R \gtrsim 0.25$~au, moments are taken from the Solar Probe Cup (SPC; \citealt{kasper_solar_2016}), while for smaller $R$ the Solar Probe Analyzer (SPAN), also part of SWEAP, is used.

Several preprocessing steps are applied to ensure data integrity. Following SWEAP instrument team recommendations, 10-minute rolling medians of the SPAN-i proton density are computed, and intervals with abrupt density drops exceeding $50\%$ are discarded. Because of partial obstruction by the spacecraft’s heat shield, we further require that the azimuthal flow angle in the instrument frame remain below approximately $165^\circ$.

Rather than directly adopting proton number densities $n_p$ from the plasma moments, we use electron densities from quasi-thermal noise (QTN) spectroscopy performed by the FIELDS instrument \citep{Moncuquet_2020}. Assuming charge neutrality and a 4\% alpha-particle abundance, we convert $n_e$ to $n_p$ via $n_p = \frac{n_e}{1.08}$ \citep[eg.,][]{2021LiuAA}. Intervals in which the SPAN-i proton density differs from the corresponding QTN estimate by more than an order of magnitude are excluded.

The parallel electron heat flux, $q_{\parallel,e}$, is derived from SPAN-Electrons (SPAN-E) measurements \citep{Whittlesey_2020}, following the methodology of \citet{Halekas_2020, Halekas_2021}. Electron temperatures from both the SPAN-E instrument and QTN method were evaluated \citep[see also][]{2020APJSMaksimovic,2023AGUFMLiu,2023LiuAA}. QTN-derived electron temperature profiles display a systematic discrepancy from the ones derived from SPAN-E. Specifically, (1) both the thermal and total electron temperatures derived from the QTN method show  steeper radial profiles than the thermal one derived from SPAN-E; and (2) the SPAN-E-derived electron thermal temperature for most of the time presents a much more pronounced anti-correlation with the proton bulk speed than the QTN ones. This systematic discrepancy among them for different solar wind populations needs further investigation and beyond the scope of the current work. In this work, we adopt SPAN-E-derived thermal electron temperature, applying the methodology and selection criteria of \citet{Halekas_2020, Halekas_2021}.

All measurements at heliocentric distances $R \geq 100,R_{\odot}$ are excluded due to degraded instrument performance. Velocity fluctuation amplitudes are estimated using the surrogate expression
\begin{equation}
\delta V^2 \equiv \delta V_R^2 + 2,\delta V_N^2,
\end{equation}
where $\delta V_R^2$ and $\delta V_N^2$ denote the variances of the radial and normal velocity components, respectively. This form is chosen to mitigate biases arising from incomplete sampling of the tangential component, $\delta V_T^2$, which is under-resolved in the $+T$ direction due to SPAN’s limited field of view. While the conventional three-component formulation yields comparable results, the surrogate definition is adopted here to enable direct comparison with the methodology of \citet{Ruffolo2020}.

\vspace{0.5em}
\subsection{Solar Orbiter (\citetalias{muller_solar_2020})}\label{Appendix_Data:SolO}

Magnetic-field and particle measurements from the Solar Orbiter (\citetalias{muller_solar_2020}) mission were analyzed over the interval 1~July~2020 to 1~March~2023. The magnetic-field measurements were provided by the Magnetometer (MAG) instrument \citep{horbury_solar_2020}, and the ion bulk properties were obtained from the Proton and Alpha Particle Sensor (SWA-PAS) of the Solar Wind Analyser (SWA) suite \citep{owen_solar_2020}. In accordance with the procedure adopted for the \citetalias{fox_solar_2016} dataset, the proton number density $n_p$ was inferred from electron number densities determined via QTN measurements.

\vspace{0.5em}
\subsection{Data Processing}\label{Appendix_methods:Data_processing}

The dataset was partitioned into contiguous segments of duration $D = 8\,\mathrm{h}$, with successive segments overlapping by $4\,\mathrm{h}$. Only segments containing  quasi-thermal noise (QTN) measurements were retained. A segment was discarded if the fraction of missing samples exceeded $2\%$ for the magnetic-field series, $10\%$ for the velocity series, or $50\%$ for the QTN-derived quantities. Plasma-moment time series were preprocessed with a Hampel filter \citep{davies_identification_1993}, applied over a $300$-point sliding window, to remove outliers exceeding three local standard deviations.

All retained segments were manually inspected to identify intervals contaminated by heliospheric current sheet (HCS) crossings, coronal mass ejections (CMEs), or other transients. For each identified event, a temporal buffer of $10\,\mathrm{min}$ was imposed on either side, and the buffered region was excised from the dataset. Gaps in the remaining segments were filled by linear interpolation only when permitted by the exclusion criteria described below.

Prior to calculating derived physical quantities, each time series was re-examined to identify individual data gaps. Sub-segments were excluded if any single gap exceeded $30\,\mathrm{s}$ in the magnetic-field measurements, $1\,\mathrm{min}$ in the velocity measurements, or $5\,\mathrm{min}$ in the QTN-derived proton density. After computation of the target quantities, additional buffers equal to $20\%$ of the respective gap thresholds were applied before and after each identified gap, and the corresponding data points were removed to suppress edge-related artefacts in subsequent analyses.

\section{Methods} \label{Appendix:methods}
\vspace{0.5em}

% To suppress extreme outliers in speed we retain only those points whose $U_i$ fall between the $\mathrm{low\_pct}$ and $\mathrm{high\_pct}$ percentiles of the full sample (for example $0.1$ and $99.9$\,\%). 

\subsection{Radius-Dependent Speed Quantiles}\label{Appendix:methods_quantiles}

We define radius-dependent speed quantiles to classify each solar-wind observation into a speed quartile that accounts for the radial variation of the bulk flow.

Given a dataset of radial distances and bulk speeds 
$\{(R_i,\,U_i)\}_{i=1}^{N}$, the heliocentric distance range 
$[R_{\min}, R_{\max}]$ is divided into $N_x$ contiguous bins of equal logarithmic width
\begin{equation}
\Delta = \frac{\log_{10} R_{\max} - \log_{10} R_{\min}}{N_x},
\end{equation}
so that the $k$-th bin is defined as
\begin{equation}
B_k = \left[10^{\log_{10} R_{\min} + k\Delta},\; 10^{\log_{10} R_{\min} + (k+1)\Delta}\right],
\qquad k = 0,\dots,N_x-1.
\end{equation}

The geometric midpoint of bin $B_k$ is denoted
\begin{equation}
\bar{R}_k = \sqrt{R_{\min,k} R_{\max,k}},
\end{equation}
and $n_k$ denotes the number of data points in that bin.
For $N_y$ quantile levels, the corresponding probability levels are
\begin{equation}
Q_j = \frac{j}{N_y}, \qquad j = 0,\dots,N_y.
\end{equation}
In each bin, we compute the empirical quantile of the speed distribution at level $Q_j$:
\begin{equation}
y_{k,j} = \operatorname{Quantile}\left(\{U_i : R_i \in B_k\},\, Q_j\right),
\end{equation}
and assign to bin $B_k$ the weight
\begin{equation}
w_k = \sqrt{n_k}.
\end{equation}

\vspace{0.5em}
For each quantile level $j$, the dependence $y_{k,j}(\bar{R}_k)$ is fitted using two parametric forms:

\noindent
\textbf{(i) Parker-wind model:}
\begin{equation}
\hat{U}_{\mathrm{P}}(R; C, R_c) 
= C \,\sqrt{-W\!\left(-\exp[-(J(R, R_c) + 1)]\right)},
\end{equation}
where $W$ is the Lambert-$W$ function and
\begin{equation}
J(R, R_c) = 4 \left[\ln\left(\frac{R}{R_c}\right) + \frac{R_c}{R} - 1\right].
\end{equation}
The parameters $(C, R_c)$ are obtained by minimizing the weighted sum of squared residuals
\begin{equation}
\Phi_{\mathrm{P}}(C, R_c) 
= \sum_{k=0}^{N_x - 1} w_k \left[\hat{U}_{\mathrm{P}}(\bar{R}_k; C, R_c) - y_{k,j}\right]^2,
\end{equation}
subject to $C \ge 0$ and $R_c \ge 0$.

\noindent
\textbf{(ii) Empirical exponential model:}
\begin{equation}
\hat{U}_{\mathrm{E}}(R; U_\infty, R_1, a) 
= U_\infty \left[1 - \exp\left(-\left(\frac{R}{R_1}\right)^a\right)\right],
\end{equation}
with $(U_\infty, R_1, a)$ minimizing
\begin{equation}
\Phi_{\mathrm{E}}(U_\infty, R_1, a) 
= \sum_{k=0}^{N_x - 1} w_k \left[\hat{U}_{\mathrm{E}}(\bar{R}_k; U_\infty, R_1, a) - y_{k,j}\right]^2,
\end{equation}
subject to $U_\infty \ge 0$, $R_1 \ge 0$, and $a \ge 0$.

\vspace{0.5em}
Once all quantile edges $\{\hat{U}_j(R)\}_{j=0}^{N_y}$ are obtained, 
an observation $(R^\ast, U^\ast)$ is assigned to quantile bin $\kappa$ via
\begin{equation}
\kappa = \min\left\{ i \in \{1,\dots,N_y\} : U^\ast < \hat{U}_i(R^\ast) \right\},
\end{equation}
so that
\begin{equation}
\hat{U}_{\kappa - 1}(R^\ast) \le U^\ast < \hat{U}_\kappa(R^\ast).
\end{equation}

\subsection{Five-Point Increments and Els\"asser Variables}\label{Appendix_methods:5pt_Increments}

To estimate scale-dependent variances in the velocity and magnetic fields, we employ the five-point increment method of \citet{2019ApJ...874...75C}.  
For component $i \in \{R,\,T,\,N\}$ in RTN coordinates \citep{FRANZ2002217}, the increment at scale $\ell$ is
\begin{equation}
\delta \phi_{i}(r,\ell) 
= \frac{1}{\sqrt{35}}\sum_{n=-2}^{2} c_n\,\phi_{i}(r+n\ell),
\end{equation}
with $c_n = [1,\,-4,\,6,\,-4,\,1]$.
The increment scale is
\begin{equation}
\ell = \lvert \tau \cdot \boldsymbol{U}_T \rvert,
\end{equation}
where the effective velocity
\begin{equation}
\boldsymbol{U}_{T} 
= \boldsymbol{U} - \boldsymbol{V}_{sc} + \mathrm{sgn}(B^{r}_{0})\,\boldsymbol{V}_{A}
\end{equation}

combines the solar wind speed ($\boldsymbol{U}$), the spacecraft velocity ($\boldsymbol{V}_{sc}$), and the Alfv\'en speed ($\boldsymbol{V}_{A}$)-a modified form of Taylor’s frozen-in-flow hypothesis \citep{taylor_spectrum_1938}. Here, $B_{r}^{0}$ denotes the average radial magnetic field computed within a sliding window, used to determine the polarity of the background magnetic field.

This formulation accounts for both wave propagation and the spacecraft’s motion relative to the Sun, thereby defining the effective frame in which plasma fluctuations are approximately “frozen” \citep{Klein_2015}.

A local, scale-dependent mean field $\boldsymbol{\phi}_{\ell}(r)$ is computed via
\begin{equation}
\boldsymbol{\phi}_{\ell}(r) 
= \frac{1}{16}\sum_{n=-2}^{2} d_n\,\boldsymbol{\phi}(r+n\ell),
\end{equation}
with $d_n = [1,\,4,\,6,\,4,\,1]$.

The perpendicular increment is
\begin{equation}
\delta\boldsymbol{\phi}_{\perp}(r,\ell) 
= \delta\boldsymbol{\phi}(r,\ell) 
- \delta\boldsymbol{\phi}_{||}(r,\ell),
\end{equation}
where
\begin{equation}
\delta\boldsymbol{\phi}_{||}(r,\ell) 
= \left[\delta\boldsymbol{\phi}(r,\ell)\cdot \hat{\boldsymbol{\phi}}_{\ell}(r)\right]
  \hat{\boldsymbol{\phi}}_{\ell}(r),
\quad
\hat{\boldsymbol{\phi}}_{\ell}(r) 
= \frac{\boldsymbol{\phi}_{\ell}(r)}{\lvert \boldsymbol{\phi}_{\ell}(r) \rvert}.
\end{equation}

Using the perpendicular components of the velocity and magnetic-field increments, expressed in Alfv\'en units, we define the Els\"asser increments for outward ($o$) and inward ($i$) modes as
\begin{equation}
\delta \boldsymbol{z}_{\perp}^{o,i} 
= \delta\boldsymbol{v}_{\perp} 
\mp \operatorname{sgn}\!\left(B^{r}_{\ell}\right)\,\delta \boldsymbol{b}_{\perp},
\end{equation}
where $B^{r}_{\ell}$ is the radial component of the local mean magnetic field $\boldsymbol{B}_{\ell}$.  
The sign convention ensures that $\boldsymbol{z}_{o}$ corresponds to antisunward and $\boldsymbol{z}_{i}$ to sunward propagating fluctuations.

\vspace{0.5em}
\subsection{Estimating Power Spectral Densities of the Els\"asser Fields via the Continuous Wavelet Transform (CWT)}\label{Appendix_methods:wavelet_analysis}

Let $X_i(t_n)$ denote the $i$th component of a vector field $\mathbf{X}(t)$, sampled uniformly at cadence $\delta t = 1/f_{\mathrm{s}}$, where $f_{\mathrm{s}}$ is the sampling frequency and $n = 0, 1, \ldots, N-1$. The continuous wavelet transform (CWT) of $X_i$ at time $t_n$ and scale $s$ is defined as
\begin{equation}
\widetilde{X}_{i}(s, t_n) 
= 
\sum_{j=0}^{N-1} 
X_{i}(t_j)\,\psi^{*}\!\left(\frac{t_j - t_n}{s}\right),
\end{equation}
where $\psi(t)$ is the Morlet wavelet \citep{Torrence_Compo_1998}, given by
\begin{equation}
\psi(t) 
= 
\pi^{-1/4} 
\left(
   e^{i\,\omega_{0}\,t} 
   - 
   e^{-\omega_{0}^{2}/2}
\right)\,
e^{-t^{2}/2},
\end{equation}
with $\omega_0 = 6$ controlling the balance between temporal and spectral resolution. The scale $s$ is related to the wavelet center frequency $f_{\mathrm{sc}}$ via
\begin{equation}
f_{\mathrm{sc}} 
= 
\frac{\omega_{0} + \sqrt{2 + \omega_{0}^{2}}}{4\pi s}.
\end{equation}

We define the local wavelet power spectral tensor as
\begin{equation}
P_{ij}(f_{\mathrm{sc}}, t_n) 
= 
\widetilde{X}_{i}(f_{\mathrm{sc}}, t_n)\,\widetilde{X}_{j}^{*}(f_{\mathrm{sc}}, t_n),
\end{equation}
and estimate the trace power spectral density (PSD) by time-averaging its diagonal components:
\begin{equation}
E(f_{\mathrm{sc}}) 
= 
\frac{2\delta t}{N}\,\sum_{i=1}^{3} 
\sum_{n=0}^{N-1} 
P_{ii}(f_{\mathrm{sc}}, t_n),
\end{equation}
where $N$ is the number of time samples in the interval.

\medskip
\noindent
To characterize the spectral properties of counter-propagating Alfv\'enic fluctuations, we apply the wavelet-based procedure described above to the Els\"asser variables $\mathbf{Z}_{\mathrm{o}}(t)$ and $\mathbf{Z}_{\mathrm{i}}(t)$, defined as
\begin{equation}
\mathbf{Z}_{\mathrm{o},\mathrm{i}} = \mathbf{U} \,\pm\, \mathrm{sgn}(B^{r}_{0})\, \frac{\mathbf{B}}{\sqrt{\mu_0 \rho}}.
\end{equation}
 We emphasize that applying the wavelet transform directly to the full time series $\mathbf{Z}_{\mathrm{o},\mathrm{i}}(t)$ or to the fluctuation fields $\delta \mathbf{Z}_{\mathrm{o},\mathrm{i}} = \mathbf{Z}_{\mathrm{o},\mathrm{i}} - \mathbf{Z}^{0}_{\mathrm{o},\mathrm{i}}$ yields identical power spectral densities.

For each vector component $Z_{(\mathrm{o},\mathrm{i}),j}(t)$, we first compute the corresponding wavelet coefficients $\widetilde{Z}_{(\mathrm{o},\mathrm{i}),j}(f_{\mathrm{sc}}, t_n)$. These are used to construct the local wavelet power spectral tensor:
\begin{equation}
P_{(\mathrm{o},\mathrm{i}),jk}(f_{\mathrm{sc}}, t_n) = \widetilde{Z}_{(\mathrm{o},\mathrm{i}),j}(f_{\mathrm{sc}}, t_n)\,\widetilde{Z}^{*}_{(\mathrm{o},\mathrm{i}),k}(f_{\mathrm{sc}}, t_n).
\end{equation}
The trace power spectral density for each Els\"asser field is then obtained by averaging the diagonal elements of the spectral tensor over all time samples:
\begin{equation}\label{eq:Elssaser_spectra_definition}
E_{(\mathrm{o},\mathrm{i})}(f_{\mathrm{sc}}) 
= 
\frac{2 \delta t}{N} \sum_{j=1}^{3} \sum_{n=0}^{N-1} P_{(\mathrm{o},\mathrm{i}),jj}(f_{\mathrm{sc}}, t_n),
\end{equation}
which provides the total spectral energy density of outward and inward propagating fluctuations at each spacecraft-frame frequency $f_{\mathrm{sc}}$.

\vspace{0.5em}
\subsection{Estimating wave-action flux spectra}\label{Appendix_methods:wave_action_flux_spectra}

For each 8-hour interval ($D = 8~\mathrm{h}$; see Appendix~\ref{Appendix_Data:PSP}), the wavelet coefficients $\widetilde{Z}_{(\mathrm{o},\mathrm{i}),j}(f_{\mathrm{sc}}, t_n)$ are computed (Appendix~\ref{Appendix_methods:wavelet_analysis}). To minimize edge effects associated with the cone of influence (COI), the first and last 30 minutes of each interval are excluded. The remaining portions are divided into rolling windows of duration $d = 10$, $30$, and $60~\mathrm{min}$, advanced in 10-second steps. Within each window, we compute the frequency-compensated wavelet spectra $E_{o(i)} f_{\mathrm{sc}}$ (cf. Equation~\ref{eq:Elssaser_spectra_definition}), which are then recast into wave-action flux spectra according to
\begin{equation}\label{eq:wave_action_flux_spectra_definition}
E_{S^{o(i)}}(f_{\mathrm{sc}}) 
= \tilde{M}_{f} \, \mathcal{G}^{2}_{o(i)},
\end{equation}
where $\tilde{M}_{f} = \tilde{\rho}\,\tilde{U}\,\tilde{R}^{2}$ and $\mathcal{G}^{2}_{o(i)} = E_{o(i)} f_{\mathrm{sc}} \left(1 \pm \tilde{n}^{1/2}\right)\tilde{n}^{-1/4}$. This yields the power spectra of the wave-action flux variables, following Equations~\ref{eq:HO_variables} and~\ref{eq:S_p_m_definition}. Here, $\tilde{n}$, $\tilde{\rho}$, $\tilde{U}$, and $\tilde{R}$ denote the mean values within each window.

In addition, for each sub-interval, we extract frequency cuts of the total wave-action flux spectra,
\begin{equation}\label{eq:total_wave_action_flux_spectra_definition}
\mathcal{E}_{S^{\ast}} = (E_{S^{o}} - E_{S^{i}})f_{\mathrm{sc}},
\end{equation}
omitting the $f_{\mathrm{sc}}$ compensation to better illustrate the raw frequency dependence. Cuts at several characteristic frequencies are used to examine the $M_a$-dependence of $\mathcal{E}_{S^{\ast}}$. For each frequency, the median values of $\mathcal{E}_{S^{\ast}}$ are computed within $\mathcal{N} = 100$ bins, spanning the range $M_a \in [0.3,\,20]$ and spaced uniformly in logarithmic scale.

\vspace{0.5em}
\subsection{Diagnostics for Alfv\'enicity}\label{Appendix_methods:cross_helicity}

To assess the degree of Alfv\'enicity in our dataset, we employ several proxies that characterize the properties of the fluctuating fields. These diagnostics help us determine whether the observed fluctuations exhibit the signatures of Alfv\'enic turbulence---namely, strong correlations between velocity and magnetic field fluctuations, energy equipartition, and negligible compressibility.
 
The Els\"asser imbalance quantifies the relative dominance of antisunward versus sunward-propagating Alfv\'en waves and is commonly measured, in the incompressible limit, using the normalized cross helicity \citep{Velli_91_waves, Velli_93}:
\begin{equation}
    \sigma^{(\perp)}_c = -2\,\text{sgn}(\boldsymbol{B}^{r}_{\ell})\,\frac{\langle \delta \boldsymbol{v}_{(\perp)} \cdot \delta \boldsymbol{b}_{(\perp)} \rangle}{\langle \delta \boldsymbol{v}_{(\perp)}^2  +  \delta \boldsymbol{b}_{(\perp)}^2 \rangle},
    \label{eq:cross_helicity}
\end{equation}
where angle brackets $\langle \cdot \rangle \equiv \langle \cdot \rangle_{d}$ denote rolling averages computed over subintervals of duration $d = 1$ minute.

Similarly, the imbalance between magnetic and kinetic energies in the fluctuations is quantified by the normalized residual energy:
\begin{equation}
    \sigma^{(\perp)}_r = \frac{\langle \delta \boldsymbol{v}_{(\perp)}^2 \rangle - \langle \delta \boldsymbol{b}_{(\perp)}^2 \rangle}{\langle \delta \boldsymbol{v}_{(\perp)}^2 \rangle + \langle \delta \boldsymbol{b}_{(\perp)}^2 \rangle}.
    \label{eq:res_energy}
\end{equation}
In a purely Alfv\'enic state, where magnetic and kinetic energies are in equipartition, $\sigma_r$ is expected to approach zero, while deviations from zero indicate an imbalance arising from non-Alfv\'enic processes or from the nonlinear evolution of turbulence.

Throughout this work, both the perpendicular increments, $\delta \boldsymbol{\xi}_{\perp}$, and the full vector increments, $\delta \boldsymbol{\xi}$, are considered in the definitions of $\sigma_c$ and $\sigma_r$. Quantities computed using the perpendicular increments are denoted by a superscript $(\perp)$, as in $\sigma_c^{(\perp)}$ and $\sigma_r^{(\perp)}$, while those computed from the full vector increments are written without this superscript. Results based on both definitions are presented and compared in the main text and in Appendix \ref{Appendix_methods:sigmas_full}.

To quantify the compressibility of the magnetic-field fluctuations, we compute the proxy
\begin{equation}
    C_{\boldsymbol{B}}
    = \frac{\langle [\,\delta|\boldsymbol{B}|\,]\rangle^2}
           {\langle |\delta\boldsymbol{B}|\rangle^2}\,.
    \label{eq:compress}
\end{equation}

An alternative proxy replaces the denominator in Equation~\eqref{eq:compress} with the parallel increment, $\delta B_{\parallel}$ \citep[see e.g.,]{Matteini_2014}.  However, for the finite-amplitude Alfv\'enic fluctuations considered here ($\delta b/B_{0}\sim1$; \citealt{Hollweg_1974, Mallet_2021}), the total field magnitude is constrained by $|\boldsymbol{B}_0+\boldsymbol{b}|^2=\text{const}$.  In this “spherical polarization” regime \citep{barnes_large-amplitude_1974, Bruno_2001, Matteini_2014}, the fluctuations are not purely transverse but include a significant parallel component, $\delta b_{\parallel}$.  Consequently, while both proxies coincide at sub-ion scales, they can differ substantially at MHD scales, with the parallel-increment form reflecting the degree of spherical polarization rather than true compressibility.

In ideal AWs-where $|\boldsymbol{B}|$ remains constant-$C_{\boldsymbol{B}}\ll1$.  Elevated values of $C_{\boldsymbol{B}}$ therefore signal departures from pure Alfv\'enic behavior. Combined with the cross-helicity, $\sigma_{c}$, and residual-energy, $\sigma_{r}$, diagnostics, $C_{\boldsymbol{B}}$ completes a robust suite of measures for assessing the Alfv\'enicity of the observed fluctuations.

\begin{figure*}
    \centering
         \includegraphics[width=0.99\textwidth]{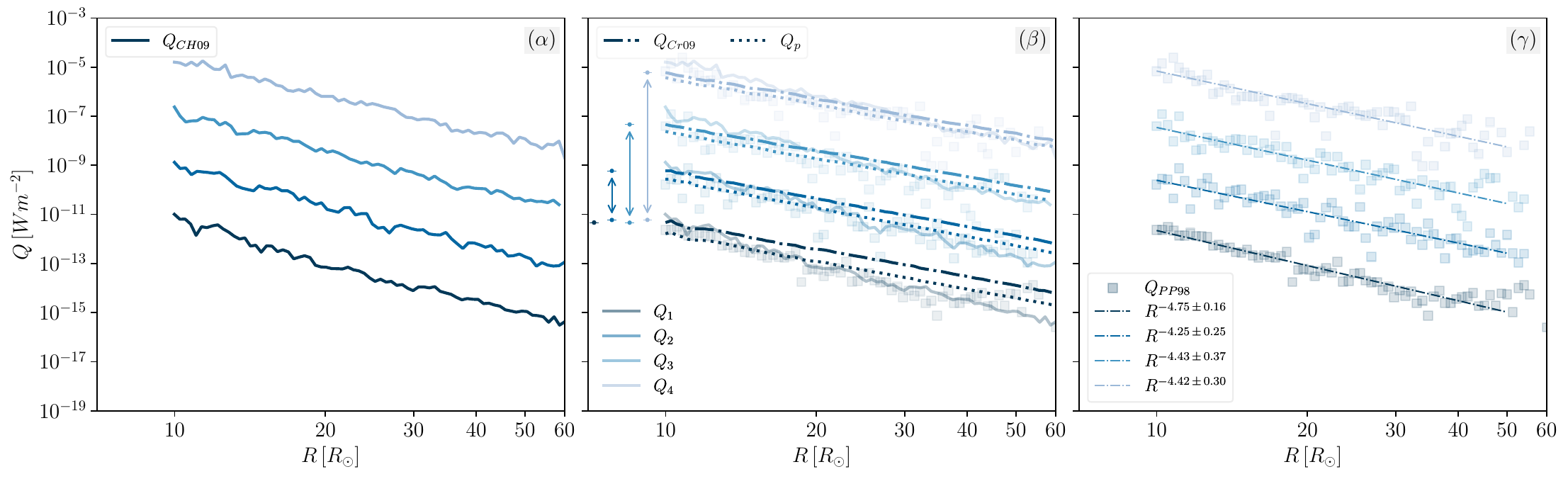}
        \caption{Volumetric heating rates, $Q$, as a function of heliocentric distance, $R$, estimated across four quartiles of the solar-wind speed distribution. Panel~($\alpha$): Estimates based on the \citetalias{CH09} formulation (solid lines). Panel~($\beta$): Results from the \citetalias{Cr09} method (dash-dotted lines), with the proton-only contribution, $Q_p$, shown as dotted lines. The \citetalias{CH09} estimates, $Q_{\mathrm{CH09}}$, are overplotted for reference (solid lines), alongside semi-transparent scatter points denoting the third-order structure-function estimates of \citetalias{PP98}. Panel~($\gamma$): $Q_{\mathrm{PP98}}$ estimates, with power-law fits (dashed lines) applied independently within each wind-speed quartile. The corresponding best-fit slopes, $\alpha$, and their uncertainties are reported in the legend. For visual clarity, all profiles are vertically offset by factors of $100^n$, where $n = 0$ (dark blue) corresponds to the slowest quartile and $n = 3$ (light blue) to the fastest.
}\label{fig:heat_rates_distance}
\end{figure*}

\vspace{0.5em}
\subsection{Evaluating Turbulent Heating Rates}\label{Appendix_methods:heating rates}

Turbulent heating rates are estimated using the \citetalias{CH09}, \citetalias{Cr09}, and \citetalias{PP98} prescriptions (Equations~\ref{eq:heating_rate_CH09}, \ref{eq:heating_rate_Cr09}, and \ref{eq:PP98_Q}), each of which requires the radial gradients of several plasma parameters. This analysis is restricted to the \citetalias{fox_solar_2016} dataset, as the electron heat-flux measurements required for evaluating $Q_e$ are not publicly available in the \citetalias{muller_solar_2020} release.

The fitted quantities are the proton and electron temperatures, $T_p$ and $T_e$ (K), the proton and electron number densities, $n_p$ and $n_e$ (m$^{-3}$), the field-aligned electron heat flux, $q_{\parallel e}$ (W\,m$^{-2}$), and the Parker spiral angle, $\Phi$ (rad). The radial coordinate is expressed in logarithmic form as
\[
x \equiv \ln\!\left(\frac{r}{\mathrm{AU}}\right),
\]
and the polynomial basis vector of degree $d$ is
\[
\mathbf{v}_d(x) \equiv \big(1,\,x,\,x^2,\,\ldots,\,x^{d}\big)^{\!\top}.
\]

For strictly positive quantities $Y\in\{T_p,T_e,n_p,n_e,q_{\parallel e}\}$, two functional forms in $x$ are considered. The first is a low-order polynomial in logarithmic space,
\begin{equation}
\ln Y(r) = \mathbf{a}_Y^{\top}\,\mathbf{v}_d(x), \qquad d\in\{1,2\},
\label{eq:poly_logfit_compact}
\end{equation}
where $\mathbf{a}_Y$ are the polynomial coefficients. The second is a continuous broken power law with a single breakpoint,
\begin{equation}
\ln Y(r) = b + k_1\,x + \Delta k\,\big(x-x_b\big)_+, \qquad (z)_+\equiv \max(z,0),
\label{eq:hinge_logfit}
\end{equation}
which in natural units becomes
\begin{equation}
Y(r) =
\begin{cases}
\displaystyle A\left(\dfrac{r}{\mathrm{AU}}\right)^{k_1}, & r< r_b,\\[8pt]
\displaystyle A\left(\dfrac{r_b}{\mathrm{AU}}\right)^{-\Delta k}\!\left(\dfrac{r}{\mathrm{AU}}\right)^{k_1+\Delta k}, & r\ge r_b,
\end{cases}
\qquad
A = e^{b}, \quad r_b = \mathrm{AU}\,e^{x_b}.
\label{eq:hinge_natural}
\end{equation}
Here $k_1$ is the slope for $r<r_b$ and $k_1+\Delta k$ is the slope for $r\ge r_b$, with $r_b$ constrained to $r_b \le r_b^{\max}$ (e.g., $40\,R_\odot$) to avoid unphysical breaks at large heliocentric distances. For the spiral angle, a polynomial of degree $d_\Phi \in \{3,4\}$ is used,
\begin{equation}
\Phi(r) = \mathbf{b}^{\top}\,\mathbf{v}_{d_\Phi}(x),
\label{eq:poly_phifit_compact}
\end{equation}
reverting to the analytic Parker-spiral form if cubic fitting is not supported by the data.

To mitigate small-scale variability and non-uniform radial sampling, all series are binned in heliocentric distance. For each bin $i$ we store the center $r_i$, the bin-averaged value $\bar{Y}_i$, and the count $N_i$. These are used as statistical weights in a one-step feasible generalized least-squares (FGLS) regression. Denoting $y_i$ as the regression target at $x_i=\ln(r_i/\mathrm{AU})$, the design matrix and response vector are
\[
V = \begin{bmatrix} \mathbf{v}_d(x_1)^\top \\ \vdots \\ \mathbf{v}_d(x_n)^\top \end{bmatrix}, 
\qquad
\mathbf{y} = \begin{bmatrix} y_1 \\ \vdots \\ y_n \end{bmatrix}.
\]
The initial weighted least-squares step uses $W_0 = \mathrm{diag}(N_i)$,
\begin{equation}
\widehat{\boldsymbol\theta} = (V^\top W_0 V)^{-1} V^\top W_0 \mathbf{y}, \qquad
\hat{\varepsilon}_i = y_i - \widehat{\boldsymbol\theta}^{\top} \mathbf{v}_d(x_i),
\end{equation}
where $\widehat{\boldsymbol\theta}$ contains the fitted coefficients and $\hat{\varepsilon}_i$ are residuals. The variance curve $\widehat{\sigma}^2(x_i)$ is estimated by fitting $\ln \hat{\varepsilon}_i^{\,2}$ to a low-order polynomial in $x$. Final heteroskedastic weights are
\begin{equation}
w_i = \frac{N_i}{\widehat{\sigma}^2(x_i)}, \qquad W = \mathrm{diag}(w_i),
\end{equation}
and the coefficient covariance is
\begin{equation}
\widehat{\mathrm{Cov}}(\widehat{\boldsymbol\theta}) = \hat{\sigma}^2 (V^\top W V)^{-1}, \quad
\hat{\sigma}^2 = \frac{1}{n} \sum_{i} w_i \left[ y_i - \widehat{\boldsymbol\theta}^{\top} \mathbf{v}_d(x_i) \right]^2.
\label{eq:sigmahat}
\end{equation}

Model selection proceeds in two stages. First, the optimal polynomial degree $d\in\{1,2\}$ in \eqref{eq:poly_logfit_compact} is determined via a step-up search. Starting from $d=1$, the candidate $d+1$ is accepted only if both the Bayesian Information Criterion (BIC),
\[
\mathrm{BIC} = k_d \ln n - 2 \,\ell_{\mathrm{max}},
\]
with $k_d = d+1$ parameters and maximum log-likelihood $\ell_{\mathrm{max}}$, decreases by at least $\Delta\mathrm{BIC} \ge 6$, and the mean predictive log-likelihood from $K=5$-fold contiguous cross-validation in $x$ improves by at least $\Delta\overline{\ell}^{\,\mathrm{CV}} \ge 0.01$. If either threshold fails, the lower degree is retained.

Second, the selected polynomial is compared against the broken power-law form \eqref{eq:hinge_logfit}--\eqref{eq:hinge_natural}, with $x_b$, $k_1$, and $\Delta k$ jointly estimated. The breakpoint is restricted to $r_b \le r_b^{\max}$ as above. The same dual thresholds, $\Delta\mathrm{BIC} \ge 6$ and $\Delta\overline{\ell}^{\,\mathrm{CV}} \ge 0.01$, are applied. The broken power law is adopted only if both are satisfied, ensuring that increased model complexity is justified by improved explanatory and predictive performance.

If $\ln Y = P_Y(x)$ is the fitted profile, the radial derivative follows from the chain rule,
\begin{equation}
\frac{\mathrm{d}Y}{\mathrm{d}r} = \frac{Y(r)}{r} P_Y'(x), \qquad
\frac{\mathrm{d}\Phi}{\mathrm{d}r} = \frac{1}{r} P_{\Phi}'(x),
\label{eq:chainrule_radial}
\end{equation}
where for the broken power law $P_Y'(x)$ is $k_1$ for $r<r_b$ and $k_1+\Delta k$ for $r\ge r_b$.

These derivatives, together with the fitted $Y(r)$, are substituted into the CH09, Cr09, and PP98 formulae to obtain $Q_{\mathrm{CH09}}$, $Q_{\mathrm{Cr09}}$, and $Q_{\mathrm{PP98}}$. The resulting profiles are then binned by both $M_a$ and heliocentric distance $R$.

The radial dependence of $Q$ is shown in Figure~\ref{fig:heat_rates_distance}, where each panel corresponds to one of the three estimation methods. Panel~($\alpha$) displays the \citetalias{CH09} estimates (solid lines); panel~($\beta$) shows the \citetalias{Cr09} results (dash-dotted lines), with proton-only contributions $Q_p$ as dotted lines and $Q_{\mathrm{CH09}}$ overplotted (solid) for comparison. Panel~($\gamma$) presents $Q_{\mathrm{PP98}}$ values (scatter), along with power-law fits (dashed lines) for each solar-wind speed quartile. The corresponding best-fit slopes, $\alpha$, and their uncertainties are reported in the legend. For clarity, all profiles are vertically offset by factors of $100^n$, where $n = 0$ (dark blue) corresponds to the slowest quartile and $n = 3$ (light blue) to the fastest.

\begin{figure*}
     \centering
        \includegraphics[width=0.99\textwidth]{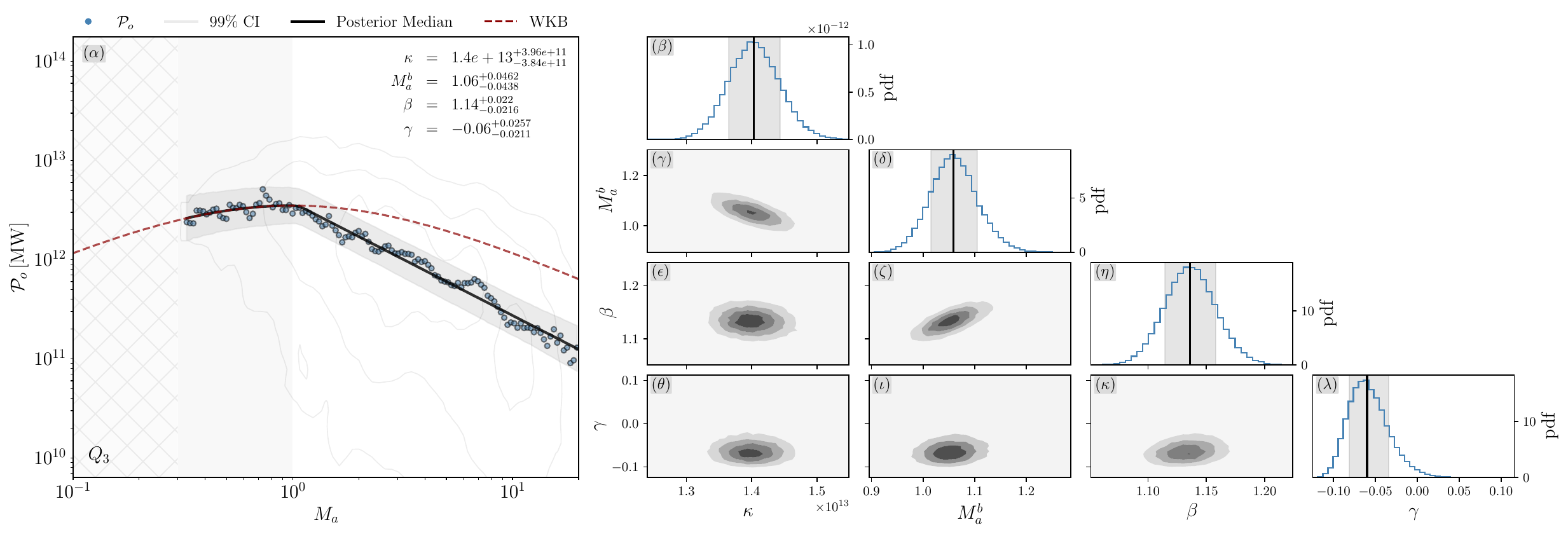}
        \caption{ (\(\alpha\)) Outward power \(\mathcal{P}_{o}\) versus \(M_a\) at fixed increment \(\tilde{\ell}=1024\,d_i\). The dashed curve is the WKB expectation \(\propto M_a/(1+M_a)^2\) fitted only over the sub-Alfv\'enic range (\(M_a<1\)). The solid black curve and gray band show the posterior median and \(99\%\) credible region of the Bayesian piecewise model of Appendix~\ref{Appendix_methods:bayesian_inference}, with a break at \(M_a^{b}\) and a non-WKB decay \(\propto (M_a/M_a^{b})^{-\beta}\) for \(M_a\ge M_a^{b}\). Data with anomalously low mass flux \((M_f<Q_{0.01}(M_f))\) are excluded; background contours denote the point density in the \((M_a,\mathcal{P}_{o})\) plane at \([0.5\sigma,1\sigma,2\sigma,3\sigma]\). (\(\beta\)–\(\iota\)) Corner plot of the posterior for \((\kappa, M_a^{b}, \beta, \gamma)\); histograms show one-dimensional marginals (vertical lines: medians), and filled contours show joint posteriors at the same probability levels. The inference favors a break just above unity and a super-Alfv\'enic decay close to a unit-slope power law, with weak \(M_a\)-dependence of the intrinsic-scatter term (\(\gamma\approx 0\)). }\label{fig:appendix_power_bayes}
\end{figure*}

\subsection{Bayesian Inference for Quantifying Departures from the WKB Profile}\label{Appendix_methods:bayesian_inference}

This appendix presents a Bayesian inference framework developed to infer the critical Alfv\'en-Mach number, $M_{a}^{b}$, at which the fluctuating energy profiles begin to deviate from the WKB prediction.

The outward-directed power, denoted $\mu(M_a; k, M_a^{b}, \beta)$, is modeled as a piecewise function of the Alfv\'en-Mach number $M_a$. The profile transitions from a WKB-like form at low $M_a$ to a power-law decay characteristic of non-WKB behavior at larger values of $M_a$. The model is specified as follows:

\begin{equation}
\mu(M_a; k, M_a^{b}, \beta) =
\begin{cases}
\displaystyle k\,\frac{M_a}{(1 + M_a)^2}, & M_a < M_a^{b}, \\[8pt]
\displaystyle k\,\frac{M_a^{b}}{(1 + M_a^{b})^2}
\left( \frac{M_a}{M_a^{b}} \right)^{-\beta}, & M_a \geq M_a^{b}.
\end{cases}
\end{equation}

In this formulation, $k$ serves as a normalization constant that sets the overall amplitude of the profile. The term $M_a / (1 + M_a)^2$ represents the expected behavior under the WKB approximation. The transition point $M_a^{b}$ marks the onset of a deviation from WKB scaling, beyond which the profile decays as a power law with exponent $-\beta$. The amplitude of the post-WKB branch is matched to the pre-transition branch to ensure continuity at $M_a = M_a^{b}$.

The observational data comprise binned ensemble averages $y_i$ of $\mathcal{P}_{o}$, corresponding to the bin-centered Alfv\'en-Mach numbers $M_{a,i}$. To accommodate measurement uncertainties and potential departures from the deterministic model, a likelihood function is constructed. The raw values are transformed to a logarithmic scale, $\ell_i = \ln y_i$, which serves to stabilize the variance and renders additive error models more appropriate, particularly in the context of strictly positive observables that span multiple orders of magnitude. The logarithmic values $\ell_i$ are then assumed to be drawn from a Student-$t$ distribution:

\begin{equation}
\ell_i \sim \text{Student-}t\!\left(\nu,\; \log\mu(M_{a,i}; k, M_a^{b}, \beta),\; \sigma_{\text{eff},i}\right).
\end{equation}

The Student-$t$ distribution is chosen for its heavier tails compared to the Normal distribution, which makes the inference more resilient to occasional outliers. The parameter $\nu$ represents the degrees of freedom of the Student-$t$ distribution; as $\nu \to \infty$, the Student-$t$ distribution approaches a Gaussian distribution.  The scale parameter, $\sigma_{\text{eff},i}$, is the effective standard deviation for the $i$-th bin, which accounts for heteroscedasticity and is defined as:

\begin{equation}
\sigma_{\text{eff},i}^{2}=
\frac{\sigma_{0}^{2}\,\mu(M_{a,i}; k, M_a^{b}, \beta)^{2\gamma}}{n_{i}}
+
\left(\frac{\sigma_{i}}{y_{i}\sqrt{n_i}}\right)^{2}.
\end{equation}

This effective variance has two components. The first term, $\sigma_{0}^{2} \mu(M_{a,i})^{2\gamma} / n_{i}$, models the intrinsic scatter. Here, $\sigma_0$ is a baseline scatter, $\mu(M_{a,i})^{2\gamma}$ allows the scatter to scale with the mean prediction (with $\gamma$ controlling the power of this scaling), and $n_i$ is the number of raw data points or counts in bin $i$, down-weighting the scatter for bins with more data. The second term, $(\sigma_{i} / (y_{i}\sqrt{n_i}))^2$, is the squared standard error of the log-mean, derived from the reported mean $y_i$ and its standard deviation $\sigma_i$ for the bin, assuming the original data within the bin are approximately log-normally distributed. This formulation ensures that both intrinsic variability and measurement precision contribute to the overall uncertainty model for each bin.

\section*{Prior Distributions and Inference Details}

Weakly informative priors are assigned to all model parameters---$k$, $M_a^{b}$, $\beta$, $\sigma_0$, $\gamma$, and $\nu$---as listed in Table~\ref{tab:priors}. These priors impose minimal physical and empirical constraints, preserving flexibility while preventing unphysical regions of parameter space.

\begin{figure*}
        \centering
         \includegraphics[width=1\textwidth]{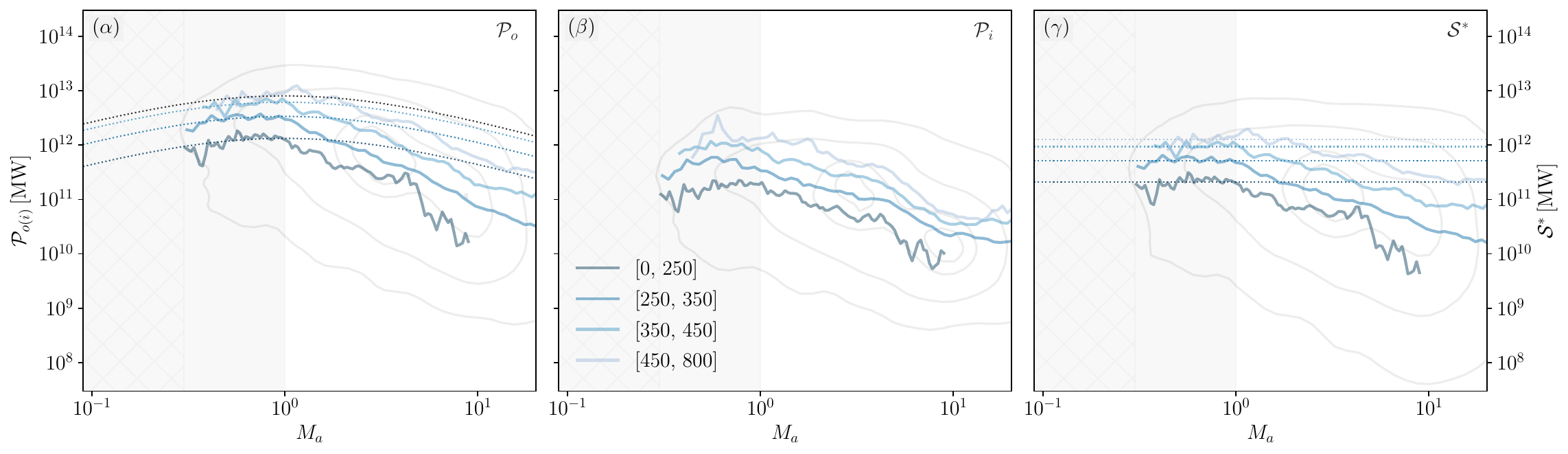}
        \caption{ Power in the ($\alpha$) outward, $\mathcal{P}_{o}$, ($\beta$) inward, $\mathcal{P}_{i}$, fluctuations, and ($\gamma$) the total wave-action flux, $\mathcal{S}^\ast$, plotted against $M_a$. All quantities are evaluated at a fixed increment scale $\tilde{\ell} = 1024\,d_i$, corresponding to spacecraft-frame frequencies within the inertial range. Each curve represents the mean within logarithmically spaced $M_a$ bins, evaluated separately for four solar wind speed intervals $ U \in \left\{ [0, 250],\ [250, 350],\ [350, 450],\ [450, 800] \right\}~\mathrm{km\,s^{-1}}.$ In panel ($\alpha$), dashed lines denote best-fit WKB profiles, $\propto M_a/(M_a + 1)^2$, fitted over the sub-Alfv\'enic range ($M_a < 1$). In panel ($\gamma$), they indicate the median of $\mathcal{S}^\ast$ within the same regime. Data points with anomalously low mass flux ($M_f < Q_{0.01}(M_f)$) are excluded. Contours reflect the density of points in the ($M_a$, $ \mathcal{P}_{o(i)} (\mathcal{S}^\ast)$) plane at levels $[0.5\sigma, 1\sigma, 2\sigma, 3\sigma]$.}\label{fig:Power_WKB_U}
\end{figure*}

\begin{table}[h!]
\centering
\begin{tabular}{llll}
\hline
Parameter & Description & Prior Distribution & Hyperparameters \\
\hline
$k$ & Scale factor (vertical amplitude) & $\log k \sim \mathcal{N}(\mu_k, s_k^2)$  & $\mu_k = \log y_{90}$, $s_k = \text{IQR}_{\log y} / 1.35$ \\
$M_a^{b}$ & Transition Alfv\'en-Mach number (breakpoint) & $\log M_a^{b} \sim \mathcal{U}(a, b)$ & $a = \log M_{a,\min}$, $b = \log M_{a,\max}$ \\
$\beta$ & Post-WKB power-law index & $\beta \sim \text{TN}_{[0,10]}(\mu_\beta, \sigma_\beta^2)$ & $\mu_\beta = 2$, $\sigma_\beta = 1.5$ \\
$\sigma_0$ & Baseline intrinsic scatter & $\sigma_0 \sim \text{Half-Normal}(s_{\sigma_0})$ & $s_{\sigma_0} = \text{MAD}_{\text{residuals}}$ \\
$\gamma$ & Scaling exponent for intrinsic scatter & $\gamma \sim \text{TN}_{[-1,\infty)}(\mu_\gamma, \sigma_\gamma^2)$ & $\mu_\gamma = 0$, $\sigma_\gamma = 0.5$ \\
$\nu$ & Degrees of freedom for Student-$t$ likelihood & $\nu \sim \text{Log-Uniform}(a_\nu, b_\nu)$ & $a_\nu = 2$, $b_\nu = 50$ \\
\hline
\end{tabular}
\caption{
\textit{ Summary of prior distributions for model parameters. $\mathcal{N}$ denotes a Normal distribution, $\mathcal{U}$ a Uniform distribution, $\text{TN}_{[L,U]}$ a Truncated Normal distribution with lower bound $L$ and upper bound $U$, and $\text{Half-Normal}(s)$ a Half-Normal distribution with scale parameter $s$. $\text{IQR}_{\log y}$ is the interquartile range of the log-transformed $y_i$ values. The scale parameter $s_{\sigma_0}$ is set to the Median Absolute Deviation (MAD) of the within-bin residuals from an initial fit.}}
\end{table}

\subsection*{ Markov Chain Monte Carlo (MCMC) Sampling with NUTS}

Samples from the joint posterior distribution over the model parameters ($k$, $M_a^{b}$, $\beta$, $\sigma_0$, $\gamma$, $\nu$) are obtained using the No-U-Turn Sampler (NUTS), an adaptive variant of Hamiltonian Monte Carlo (HMC), implemented via \citetalias{PYMC_2015}. NUTS is designed for efficient sampling from complex, high-dimensional posteriors by automatically tuning step size and trajectory length. The sampling procedure uses 16 independent Markov chains. Each chain undergoes a 4,000-iteration warm-up phase, during which adaptation is performed; these iterations are discarded. After warm-up, each chain generates 4,000 retained draws, yielding 16,000 posterior samples in total.

To promote rapid convergence, initial parameter values are set using moment-based estimates derived from the data. The sampler is configured with a target acceptance probability of 0.97 and a maximum tree depth of 10. These settings are chosen to reduce the occurrence of divergences when sampling from posteriors with sharp curvature or strong correlations.

Convergence and adequate posterior exploration are assessed using standard Markov chain diagnostics. The primary criterion is the potential scale reduction factor, $\hat{R}$ (Gelman-Rubin statistic), computed for each parameter; all chains are required to satisfy $\hat{R} \leq 1.05$. In addition, the effective sample size (ESS) is evaluated, with particular emphasis on the bulk-ESS. A minimum bulk-ESS of 400 is enforced to ensure reliable posterior estimates. Sampler divergences are also monitored; well-converged chains should exhibit few or none.

In addition to MCMC diagnostics, posterior predictive checks (PPCs) are used to evaluate the model’s capacity to reproduce the observed data. Specifically, the Pareto-smoothed importance sampling leave-one-out cross-validation (PSIS-LOO) criterion is computed, yielding an estimate of the expected log pointwise predictive density (ELPD) for new data. This metric assesses out-of-sample predictive accuracy and identifies potentially influential observations. Visual PPCs are also performed by comparing realizations from the posterior predictive distribution to the observed data.

To assess the influence of prior choices on the inferred parameters, a prior sensitivity analysis is performed. This involves perturbing the priors for $k$, $\beta$, $\sigma_0$, $\gamma$, and $\nu$ (the prior for $M_a^{b}$ is already broad) by approximately one standard deviation in directions expected to influence the posterior of $M_{a}^{b}$, followed by repeated MCMC sampling. The resulting change in the posterior median of $M_{a}^{b}$ is used as a diagnostic. Under these perturbations, the posterior median of $M_{a}^{b}$ shifts by less than 3\%, indicating that the inference is dominated by the likelihood and is robust to the specific prior assumptions.

The primary output of the Bayesian inference is the posterior distribution of the transition Alfv\'en Mach number, $p(M_{a}^{b} \mid \text{data})$, which encodes all information about $M_{a}^{b}$ conditioned on the observed turbulent profile and the specified model. Summary statistics are computed from the marginal posterior, with the posterior median reported as a point estimate. Uncertainty is characterized using credible intervals, including both the 95\% highest posterior density (HPD) interval and the central 95\% credible interval. A 95\% credible interval $[M_{a,L}, M_{a,U}]$ defines the range within which $M_{a}^{b}$ lies with posterior probability 0.95.

\begin{figure*}
     \centering
        \includegraphics[width=0.99\textwidth]{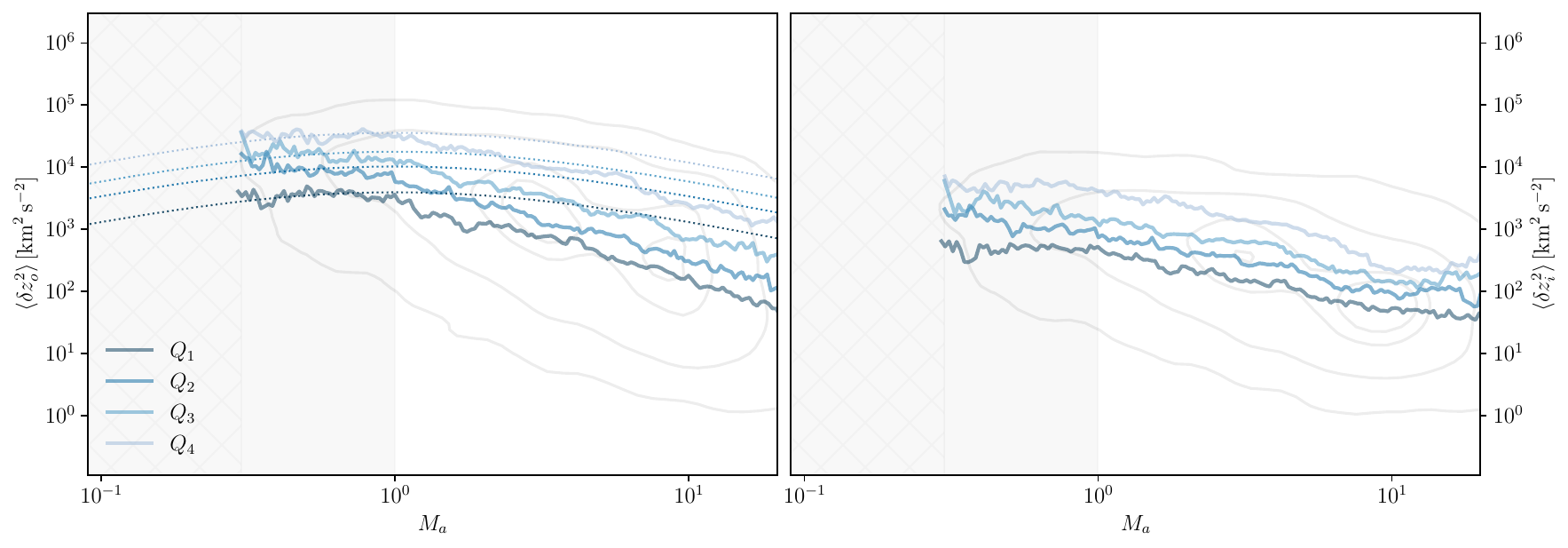}
        \caption{Variance of the ($\alpha$) outward and ($\beta$) inward propagating Els\"asser fields as a function of $M_{a}$, evaluated across different solar wind speed bins. Dashed lines in panel ($\alpha$) indicate the  best-fit WKB profile, $\propto M_a/(M_a + 1)^2$, determined over the sub-Alfv\'enic interval ($M_a < 1$) and color-coded by speed bin. Contour levels represent $[0.5\sigma, 1\sigma, 2\sigma, 3\sigma]$ intervals of the distribution}\label{fig:appendix_RMS}
\end{figure*}

\section{Some Further Results}

\subsection{Sensitivity to Binning Scheme Based on Bulk Solar Wind Speed}\label{appendix_furth_res_U_bins}

To evaluate the sensitivity of the results to the choice of $U$-based binning, the analysis is repeated using an alternative scheme in which the data are partitioned into four fixed bulk-speed intervals: 0-250~km\,s$^{-1}$, 250-350~km\,s$^{-1}$, 350-450~km\,s$^{-1}$, and 450-800~km\,s$^{-1}$. Within each bin, the dependence of $\mathcal{P}_{o(i)}$ and $\mathcal{S}^{\ast}$ on $M_a$ is re-evaluated following the same procedure described in Section~\ref{sec:Statistical Results}.

The resulting profiles, shown in Figures~\ref{fig:Power_WKB_U}$\alpha$–$\gamma$, qualitatively reproduce the trends obtained using the acceleration-based binning in the main analysis. In the sub-Alfv\'enic regime ($M_a < 1$), $\mathcal{P}_{o}$ adheres to the WKB scaling, and the wave-action flux $\mathcal{S}^\ast$ remains approximately constant across all $U$ bins. The characteristic transition near $M_a \sim 1$ and the gradual balancing of counter-propagating fluxes at higher $M_a$ are also preserved. These results indicate that, for binning schemes based on bulk solar wind speed, the behavior of fluctuations in the sub-Alfv\'enic regime is largely insensitive to whether acceleration profiles or constant-$U$ thresholds are used.

In the super-Alfv\'enic regime ($M_a > 1$), however, the two binning schemes yield systematically different scaling behavior. Specifically, the decay of $\mathcal{P}_o$ is significantly steeper under constant-$U$ binning, with a scaling index $\beta \approx -1.7$ to $-1.8$, compared to $\beta \approx -1.0$ to $-1.1$ in the acceleration-based scheme. This discrepancy indicates that, although the location of the transition away from WKB scaling is robust, the inferred post-transition scaling is sensitive to the binning methodology. The steepening observed under constant-$U$ binning may result from differences in how the $(R, M_a)$ space is sampled, which can alter the relative weighting of high-$M_a$ intervals.

\subsection{Variance as a WKB Diagnostic}\label{Appendix_methods:WKB_adherence}

Appendix~\ref{Appendix_Theory:Wave_action} reviews the theoretical basis for wave-action conservation and outlines the conditions under which the WKB approximation applies. In this framework, adherence to WKB scaling can be tested by examining the antisunward fluctuation power, $\mathcal{P}_o$, as function $M_a$. Under the assumption of constant mass flux, $M_f = \text{const}$ \citep[e.g.,][]{bavassano_radial_1982,Horbury_2001,Tenerani_2021}, Equation~\ref{eq:WKB_Power_Ma} reduces to
\begin{equation}\label{eq:variance_Ma}
\langle \delta z_{o}^{2} \rangle \propto \frac{M_a}{(M_a + 1)^2} .
\end{equation}

Figure~\ref{fig:appendix_RMS}$\alpha$ shows that the variance $\langle \delta z_{o}^{2} \rangle$ adheres to the WKB scaling given by Equation~\eqref{eq:variance_Ma} across the sub-Alfv\'enic regime, followed by a transition to power-law decay at higher $M_a$.

\begin{figure*}
     \centering
        \includegraphics[width=0.99\textwidth]{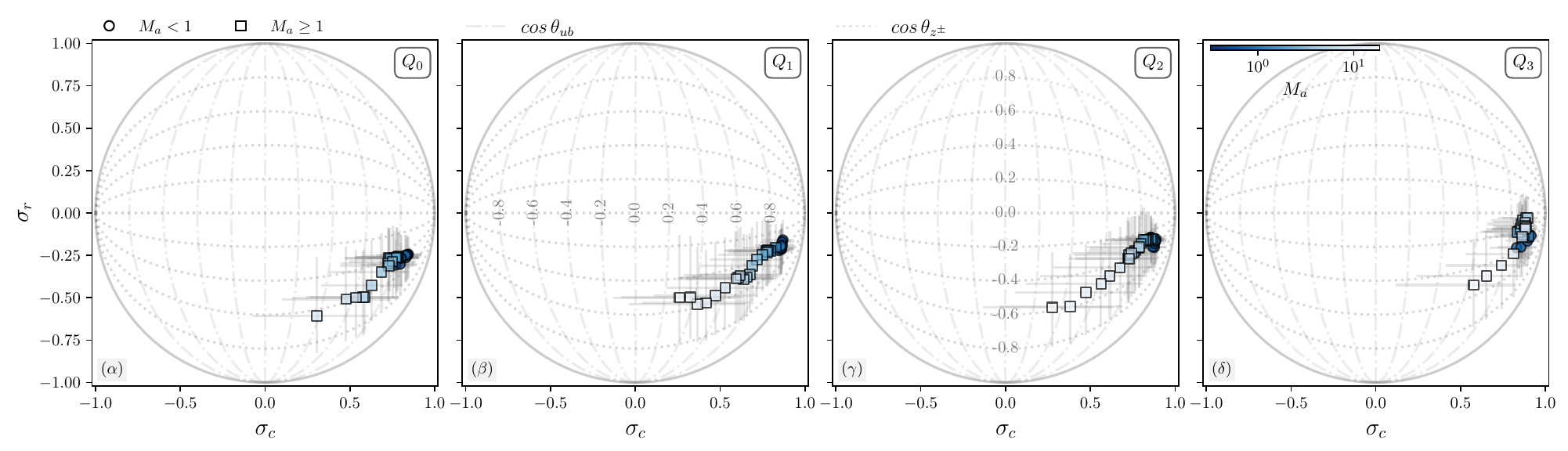}
        \caption{Parametric representation of $\sigma_r$ and $\sigma_c$ as functions of $M_a$. Each marker denotes the median within one of 30 logarithmically spaced $M_a$ bins, with horizontal and vertical error bars indicating the interquartile range (25th-75th percentiles). Circles are used for bins with $M_a \leq 1$, while squares denote bins with $M_a > 1$. Panels (a-d) correspond to the four quartiles of the solar wind acceleration profiles (see Appendix \ref{Appendix:methods}). The logarithmic color scale encodes the bin-centered value of $M_a$. Horizontal dashed gray lines indicate contours of constant Els\"asser increment alignment angle, defined by $\cos\, \theta_{z^{\pm}} = \sigma_{r}/\sqrt{1 - \sigma_{c}^{2}}$, while vertical dash-dotted gray lines mark contours of constant velocity-magnetic field increment alignment angle, defined by $\cos\, \theta_{ub} = \sigma_{c}/\sqrt{1 - \sigma_{r}^{2}}$.
}\label{fig:sigmas_full_comp}
\end{figure*}

\vspace{0.5em}
\subsection{Joint $\sigma_c -\sigma_r$ distribution: Full Increments}\label{Appendix_methods:sigmas_full}

A critical methodological consideration in this analysis is the definition of the field increment used to compute $\sigma_c$ and $\sigma_r$. Close to the Sun, where $\delta \xi_{\perp} \gg \delta \xi_{\parallel}$, the choice between the full vector increment, $\delta \bm{\xi}$, and its perpendicular component, $\delta \xi_{\perp}$, has negligible effect. At larger heliocentric distances, however, $\delta \xi_{\parallel}$ increases in amplitude relative to $\delta \xi_{\perp}$, and this distinction becomes nontrivial as $M_a(R)$ increases.

When the analysis is restricted to $\delta \xi_{\perp}$, the resulting distribution in the $\sigma_r$–$\sigma_c$ plane lies closer to the unit-circle boundary (Figure~\ref{fig:sigmas}). In contrast, using full vector increments (Figure~\ref{fig:sigmas_full_comp}) produces a distinct trajectory: although both $\sigma_c$ and $\sigma_r$ decline with increasing $M_a(R)$, the degree of Els\"asser alignment also weakens. This reduction is reflected in the systematic inward shift of points from the circle’s perimeter toward its center, relative to the perpendicular-increment case.

\bibliography{sample631}
\bibliographystyle{aasjournal}

\end{CJK*}
\end{document}